\documentclass[]{aastex701}

\usepackage{enumitem}
\usepackage{xcolor}
\usepackage{amsmath}
\usepackage{multirow}

\begin{document}

\title{Intrinsic Soft X-ray Emission from the Hot Circumgalactic Medium in IllustrisTNG}
\shorttitle{Hot-CGM X-Ray Emission in IllustrisTNG}
\shortauthors{Kim \& Kim}

\author[orcid=0009-0008-5027-8066]{Minjae Kim}
\affiliation{Department of Physics, Korea Advanced Institute of Science and Technology (KAIST), 291 Daehak-ro, Yuseong-gu, Daejeon 34141, Republic of Korea}
\email[show]{heromj27@kaist.ac.kr}

\author[orcid=0000-0003-1954-5046]{Junhan Kim}
\affiliation{Department of Physics, Korea Advanced Institute of Science and Technology (KAIST), 291 Daehak-ro, Yuseong-gu, Daejeon 34141, Republic of Korea}
\email[show]{junhan@kaist.ac.kr}

\correspondingauthor{Junhan Kim}

\begin{abstract}
We study the intrinsic 0.5--2~keV emission of the hot circumgalactic medium (CGM) in the IllustrisTNG simulation TNG100-2 at redshifts 0, 0.1, and 0.2. We calculate the emission of gas cells with $T \geq 10^6$~K in the spherical shell between $0.15R_{500}$ and $R_{500}$ and compare samples selected by their star-formation activity. X-ray luminosity increases with halo mass, although the class-dependent fits cover different mass intervals. Among galaxies with positive luminosity and star formation rate (SFR), the high-activity sample retains a positive association after a linear adjustment for halo mass. The residual association is weaker in the quiescent subset with positive luminosity and SFR. Within $10^{11.75}<M_{500}/M_\odot<10^{12.25}$, the high-activity sample has more centrally concentrated emission and a median luminosity about five times that of the quiescent sample. The two populations have similar median hot-gas masses despite their different X-ray luminosities and radial profiles. These differences are consistent with variations in the spatial distribution and thermodynamic state of the emitting gas, although the samples are not matched in halo mass. Our results suggest that soft X-ray emission probes the structure of the hot CGM as well as its gas content, offering a way to investigate how star formation and feedback shape gaseous halos.
\end{abstract}


\section{INTRODUCTION}
The circumgalactic medium (CGM) surrounds galaxies outside their interstellar medium (ISM), extending roughly to the virial radius \citep{Tumlinson2017ARA&A..55..389T}. It contains a substantial reservoir of baryons that are not accounted for by stars and the ISM \citep[e.g.,][]{Fukugita1998ApJ...503..518F, Bregman2007ARA&A..45..221B}. This multiphase gas mediates the exchange of matter and energy between galaxies and the intergalactic medium through inflows \citep{Keres2005MNRAS.363....2K} and galactic outflows \citep{Veilleux2005ARA&A..43..769V}. Observations of the CGM can therefore help constrain the supply of gas for star formation and the effects of feedback.

Efforts to understand the spatial distribution and gas content of galactic halos date back to the 1950s and 1960s \citep[e.g.,][]{Spitzer1956ApJ...124...20S,Munch1961ApJ...133...11M}. Absorption lines in background quasar spectra subsequently provided a way to probe gas too faint to detect in emission \citep{Bahcall1969ApJ...156L..63B, Bergeron1986A&A...155L...8B}. Current probes include absorption spectroscopy \citep{Tumlinson2011Sci...334..948T, Werk2014ApJ...792....8W}, fast radio bursts \citep{Prochaska2019MNRAS.485..648P, Macquart2020Natur.581..391M}, X-ray emission \citep{Forman1985ApJ...293..102F, Anderson2015MNRAS.449.3806A, Zhang2024A&A...690A.267Z}, and the thermal and kinematic Sunyaev--Zel'dovich effects \citep{Hand2012PhRvL.109d1101H, Schaan2021PhRvD.103f3513S}. Their different dependences on density, temperature, ionization state, and gas motion make them complementary, with different observational limits in halo mass and redshift \citep{Battaglia2019BAAS...51c.297B}.

Hot CGM gas can be heated by gravitational accretion shocks and by stellar or active galactic nucleus (AGN) feedback. Stable accretion shocks depend on halo mass and gas cooling, so a virialized hot atmosphere is not expected in every halo \citep{White1978MNRAS.183..341W, Birnboim2003MNRAS.345..349B}. Its X-ray emission includes both continuum radiation and metal lines, with their relative contributions depending on temperature and abundance \citep{Lovisari2021Univ....7..139L}. The thermal Sunyaev--Zel'dovich (tSZ) effect provides a complementary measurement through Compton scattering of cosmic microwave background photons by free electrons \citep{Sunyaev1970Ap&SS...7....3S,Sunyaev1972CoASP...4..173S, Sunyaev1980ARA&A..18..537S}. It measures the line-of-sight integral of electron pressure ($P_e \propto n_\mathrm{e} T$), rather than selecting gas above a fixed temperature threshold \citep{Mroczkowski2019SSRv..215...17M}. Using cosmological simulations, \citet{Junhan2022ApJ...926..179K} examined the tSZ signal and CGM thermodynamic profiles and found a strong dependence of the inferred feedback effects on aperture. This motivates a comparison of pressure-weighted and X-ray measurements over the same radial region.

For optically thin gas, the X-ray emissivity is proportional to $n_{\rm e}n_{\rm H}\Lambda(T,\{\mathcal{A}_X\})$, where $\Lambda$ is the emissivity in the chosen energy band and $\{\mathcal{A}_X\}$ denotes the elemental abundances \citep{Sarazin1986RvMP...58....1S}. At fixed composition and temperature, this weighting gives dense gas a much larger contribution than its mass alone would suggest. The low surface brightness of the outer CGM consequently makes emission near the virial radius difficult to detect around individual galaxies \citep{Anderson2011ApJ...737...22A}. The luminosity--mass relation provides a useful summary of the emission, but its interpretation requires a band-specific reference model. In the soft X-ray band, the temperature dependence of metal-line emission prevents a single bolometric self-similar slope from serving as a universal benchmark \citep{Lovisari2021Univ....7..139L}.

Given the intrinsic faintness of this emission, early X-ray studies relied mainly on pointed observations of nearby galaxies \citep[e.g.,][]{Forman1985ApJ...293..102F, Strickland2004ApJS..151..193S, Li2013MNRAS.428.2085L}, while stacking extended the measurement to larger samples \citep{Anderson2015MNRAS.449.3806A}. With eROSITA survey data, these measurements now constrain average luminosities and radial profiles into the galaxy-mass regime. For example, \citet{Zhang2024A&A...690A.268Z} measured a slope of $1.32\pm0.05$ for the luminosity--halo-mass relation from stacked data. \citet{Popesso2025A&A...704A.278P} used a different group selection and halo-mass calibration and obtained a steeper relation. These results illustrate the sensitivity of the inferred scaling relation to the halo-mass distribution and the construction of the stack. Differences in slope alone do not determine the relative importance of gravitational and feedback heating.

The inferred dependence on galaxy activity also changes with sample selection. Analyses of the eROSITA Final Equatorial‑Depth Survey (eFEDS) field reported different luminosity contrasts between star-forming and quiescent galaxies \citep{Comparat2022A&A...666A.156C, Chadayammuri2022ApJ...936L..15C}. \citet{Zhang2025A&A...693A.197Z} examined these selection issues and found similar stacked X-ray emission for the two populations in halo-mass bins with $\log_{10}(M_{200\rm m}/M_\odot)>12.5$. This differs from the comparison at fixed stellar mass, where quiescent central galaxies typically occupy more massive halos \citep{Mandelbaum2016MNRAS.457.3200M, Tinker2013ApJ...778...93T, Rodriguez-Puebla2015ApJ...799..130R}. Halo mass, activity selection, and the radial region being measured must therefore be distinguished when comparing a predicted luminosity contrast with these observations.

Simulations allow the emission to be related to the gas properties of individual halos. Using the Feedback In Realistic Environments (FIRE) simulations, \citet{Voort2016MNRAS.463.4533V} found that X-ray luminosity was closely associated with star formation in lower-mass halos, whereas more massive systems were dominated by a quasi-hydrostatic hot atmosphere. Their calculations reproduced several trends in the stacked measurements of \citet{Anderson2015MNRAS.449.3806A} without AGN feedback. That agreement applies to the quantities and halo sample studied there, and it does not establish that AGN feedback is generally unimportant. Their use of multiple simulation times increased the sampling of variability, but not the number of independent halo histories.

Activity-dependent X-ray emission has also been studied in simulations that include AGN feedback. \citet{Oppenheimer2020ApJ...893L..24O} used EAGLE and IllustrisTNG to predict stacked eROSITA measurements and related the luminosity contrast to differences in halo gas content. \citet{Truong2020MNRAS.494..549T} calculated intrinsic X-ray emission in IllustrisTNG and connected the lower luminosities of quiescent galaxies at fixed stellar mass to gas depletion and the black-hole feedback model. These studies establish the relevance of galaxy activity and gas content, but their apertures, temperature selections, and sample definitions differ from the hot, core-excised CGM considered here.

Here we examine the intrinsic 0.5--2~keV emission of gas with $T\geq10^6$~K in the spherical shell between $0.15R_{500}$ and $R_{500}$ in TNG100-2. We describe its dependence on halo mass, stellar mass, and star-formation activity using three operational activity classes. We then focus on Milky Way-scale halos and compare the high-activity star-forming and quiescent samples. We combine X-ray luminosities and radial concentration with pressure-weighted tSZ profiles and hot and all-phase gas masses, all measured within the same core-excised aperture. This joint
comparison allows us to examine how the emission relates to the gas content and radial structure of the selected populations within a restricted halo-mass interval. Our measurements concern intrinsic three-dimensional halo properties. Quantitative comparisons with observational stacks additionally require accounting for projection, source contamination, sample selection, and averaging.

This paper is organized as follows.
Section~\ref{sec:data_methods} describes the simulation, sample selection, and X-ray calculation. Section~\ref{sec:LX_mass_SFR_relation} presents the luminosity relations with mass and star formation rate. Section~\ref{sec:Star-Forming vs Quiescent galaxy in Milky Way mass galaxy} compares the radial profiles and gas masses of the Milky Way scale samples. We discuss the interpretation and comparison with earlier work in Section~\ref{sec:discussion} and summarize the results in Section~\ref{sec:conclusion}. 

We use the cosmology adopted by IllustrisTNG, with $\Omega_{\Lambda,0}=0.6911$, $\Omega_{\rm m,0}=0.3089$, $\Omega_{\rm b,0}=0.0486$, $\sigma_8=0.8159$, $n_s=0.9667$, and $h=0.6774$ \citep{planck2016A&A...594A..13P}. Here $h=H_0/(100\,{\rm km\,s^{-1}\,Mpc^{-1}})$ and $E(z)=H(z)/H_0$.

\section{DATA AND METHODS} \label{sec:data_methods}

We use the IllustrisTNG cosmological simulations \citep{Nelson2018MNRAS.475..624N,Springel2018MNRAS.475..676S,Marinacci2018MNRAS.480.5113M,Pillepich2018MNRAS.475..648P,Naiman2018MNRAS.477.1206N}. The simulations employ the moving-mesh code AREPO \citep{Springel2010MNRAS.401..791S} and include radiative cooling, star formation, stellar and black-hole feedback, and magnetic fields. Halos are identified with a friends-of-friends (FoF) algorithm \citep{Davis1985ApJ...292..371D}, and gravitationally bound subhalos are then identified with \texttt{SUBFIND} \citep{Springel2001MNRAS.328..726S}. Our analysis uses TNG100-2, the intermediate-resolution realization of the TNG100 volume. Its target baryonic mass resolution is $7.55\times10^6\,h^{-1}M_\odot$, and its dark matter particle mass is $4.04\times10^7\,h^{-1}M_\odot$. This run provides a large full-physics galaxy sample while keeping the cell-level X-ray emissivity calculation computationally tractable.

\subsection{Galaxy Sample Selection and Classification} \label{subsec:Simulation and Galaxy Sample Selection}

We select central galaxies using a procedure based on \citet{Junhan2022ApJ...926..179K}, with the redshift range and activity classes defined below. We use snapshots 99, 91, and 84, corresponding approximately to $z=0$, $0.1$, and $0.2$, respectively. These redshifts overlap those of recent X-ray stacking studies \citep[e.g.,][]{Zhang2024A&A...690A.268Z}. We select halos with $10^{11}<M_{500}/M_\odot<10^{13.5}$ and identify their central subhalos using the \texttt{GroupFirstSub} catalog field, retaining objects with \texttt{SubhaloFlag}=1. Each central subhalo must contain more than 50 gas cells and more than 50 type-4 particles. The latter count follows the simulation's particle-type convention, which includes both stars and wind-phase particles.

Using the SubLink main-progenitor branches \citep{Rodriguez-Gomez2015MNRAS.449...49R}, we find that 93.6\% and 89.1\% of the galaxies in our $z=0$ sample have progenitors that also pass our selection at $z=0.1$ and $z=0.2$, respectively. The pooled sample therefore contains repeated appearances of the same galaxy lineages. We distinguish these halo--snapshot entries from independent galaxies.

For the primary activity classification, we use the instantaneous gas-based SFR and an aperture mass estimate denoted by $M_\star$, following the general approach of previous TNG studies \citep{Truong2020MNRAS.494..549T,Oppenheimer2020ApJ...893L..24O}. Both quantities are measured within a spherical aperture of radius 30 physical kpc centered on the central subhalo, using members of its host FoF halo. Distances account for periodic boundaries and are converted to physical units at each snapshot.

The SFR is the sum of the \texttt{StarFormationRate} values of gas cells within the aperture. The mass estimate is the sum of the current masses of type-4 particles in the same aperture, without a separate stellar-formation-time selection. A fixed physical aperture need not contain the same fraction of a galaxy as an aperture defined using its stellar half-mass radius \citep{Donnari2019MNRAS.485.4817D}.

\begin{figure*}[t!]
    \centering
    \includegraphics[width=\linewidth]{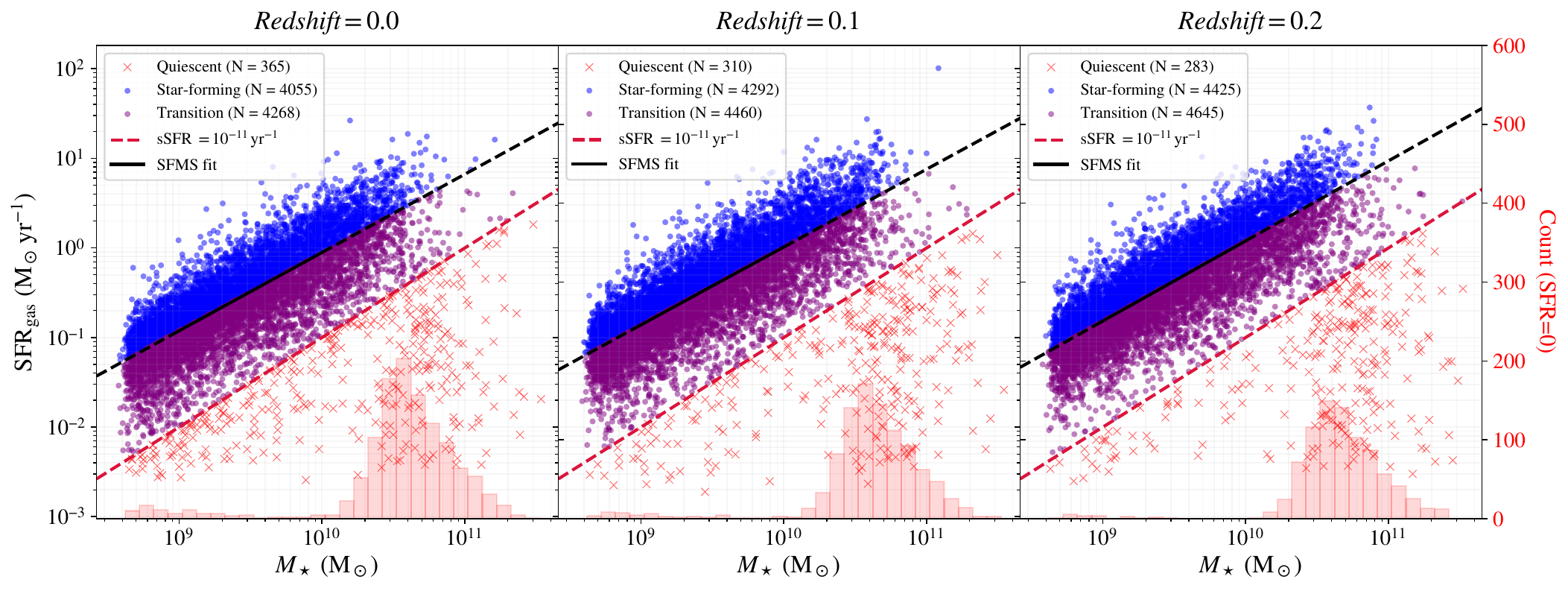}
    \caption{Star formation rate (SFR) versus stellar mass, with galaxies classified by their specific star formation rate (sSFR) and offset from the fitted star-forming main sequence. The quiescent sample (red) lies below ${\rm sSFR}=10^{-11}\,{\rm yr}^{-1}$. Among the remaining galaxies, the star-forming sample (blue) lies on or above the fitted star-forming main sequence (SFMS), and the transition sample (purple) lies below it. The SFMS is fitted separately at each snapshot as described in Section~\ref{subsec:Simulation and Galaxy Sample Selection}. Legend counts refer to objects with positive SFR. The histogram separately shows the stellar-mass distribution of quiescent galaxies with ${\rm SFR}=0$, which cannot be placed on the logarithmic SFR axis. The three panels show snapshots 99, 91, and 84 ($z=0$, 0.1, and 0.2). These are operational activity classes, and the transition label does not identify galaxies observed to be quenching.}
    \label{fig:Galaxy Classification}
\end{figure*}

Figure~\ref{fig:Galaxy Classification} shows our activity classification. We define $\mathrm{sSFR}=\mathrm{SFR}/M_\star$ and classify galaxies with $\mathrm{sSFR}<10^{-11}\,\mathrm{yr}^{-1}$ as quiescent \citep[e.g.,][]{Wetzel2013MNRAS.432..336W}.
Throughout this paper, ``quiescent'' describes this low-sSFR state, whereas ``quenching'' refers to the suppression of star formation over time.

We fit the star-forming main sequence (SFMS) separately at each snapshot using non-quiescent galaxies with $\mathrm{sSFR}\geq10^{-11}\,\mathrm{yr}^{-1}$ and
$10^9<M_\star/M_\odot<10^{10}$. For each 0.2 dex stellar-mass bin containing more than
five galaxies, we calculate the mean logarithmic stellar mass and median logarithmic SFR. We fit an unweighted straight line to these bin summaries:
\begin{equation}\label{eq:sfms}
    \log_{10}\!\left(
    \frac{\mathrm{SFR}}{M_\odot\,\mathrm{yr}^{-1}}
    \right)
    =
    \alpha\log_{10}\!\left(
    \frac{M_\star}{M_\odot}
    \right)+\beta.
\end{equation}
The fitted coefficients are $(\alpha,\beta)=(0.87,-8.79)$, $(0.87,-8.68)$, and
$(0.89,-8.86)$ at $z=0$, $0.1$, and $0.2$, respectively. We extrapolate these relations outside the fitting interval to define the boundary between the star-forming and transition samples.

Among the remaining galaxies with $\mathrm{sSFR}\geq10^{-11}\,\mathrm{yr}^{-1}$, we label those on or above the fitted SFMS as star-forming and those below it as transition galaxies. These are operational labels: the star-forming class selects the higher-activity side of the SFMS, while the transition class can include ordinary star-forming galaxies below the ridge. Membership in the latter class does not establish that a galaxy is undergoing quenching. A fixed sSFR threshold need not follow the evolving SFMS, particularly at higher redshift \citep{Donnari2021MNRAS.506.4760D}. We retain this intermediate population rather than forcing a binary star-forming/quiescent split. This three-way classification allows us to examine how the hot-CGM X-ray properties vary across galaxies with different levels of star-formation activity.

\subsection{CGM Gas Selection and X-ray Emission Modeling} \label{subsec: Gas Selection of CGM and X-ray Emission Modeling}

We define hot gas by $T\geq10^6\,\mathrm{K}$. We calculate the temperature from each cell's specific internal energy and mean molecular weight, using its hydrogen mass fraction and electron abundance. We model the intrinsic 0.5--2.0 keV emission as that
of an optically thin plasma in collisional ionization equilibrium (CIE). The emission model does not include photoionization or departures from ionization equilibrium.

For each cell, the hydrogen number density is $n_{\rm H}=X_{\rm H}\rho/m_{\rm p}$, where $X_{\rm H}$ is the hydrogen mass fraction, $\rho$ is the physical gas density, and $m_{\rm p}$ is the proton mass. The electron number density is $n_e=x_en_{\rm H}$,
where $x_e$ is the electron abundance supplied by the simulation, and the cell volume is $V=m/\rho$. The luminosity of cell $j$ is
\begin{equation}\label{eq1}
    L_{\mathrm{X},j}
    =
    n_{e,j}n_{{\rm H},j}V_j\,
    \Lambda_{\rm gas}
    \left(T_j,\{\mathcal{A}_{X,j}\}\right),
\end{equation}
where $n_en_{\rm H}V$ is the emission measure, in $\mathrm{cm}^{-3}$, and $\Lambda_{\rm gas}$ is the 0.5--2.0 keV emissivity per emission measure, in $\mathrm{erg\,cm^3\,s^{-1}}$. The elemental abundances $\mathcal{A}_X$ are defined below.

We define the CGM aperture as the spherical shell $0.15R_{500}<r\leq R_{500}$ centered on the central subhalo and select gas cells belonging to its host FoF halo. Distances are evaluated with periodic boundary conditions. Here $R_{500}$ encloses a mean total matter density equal to 500 times the critical density at the snapshot redshift. The inner cut removes the central region, but no separate satellite mask is applied. The resulting luminosity is an intrinsic three-dimensional aperture measurement, rather than a projected observational measurement.

Gas-membership choices have also been examined for tSZ profiles \citep{Junhan2022ApJ...926..179K} and galaxy X-ray emission \citep{Truong2020MNRAS.494..549T}. Their effects need not be the same for these observables, because X-ray emissivity depends on density differently
from the tSZ signal. We compared the luminosities calculated from host-FoF gas
with those obtained from all spatially selected gas cells, using the same radial aperture and temperature cut. The two selections give similar luminosities for the tested sample. We therefore use host-FoF gas to reduce the computational cost.

We calculate $\Lambda_{\rm gas}$ with the Astrophysical Plasma Emission Code (APEC; \citealt{Smith2001ApJ...556L..91S}), accessed through \texttt{PyAtomDB} version 1.2.0 \citep{Foster2020Atoms...8...49F}. For each element $X$, we define the abundance supplied to the plasma model as
\begin{equation}\label{eq2}
    \mathcal{A}_X=\frac{(n_X/n_{\rm H})_{\rm gas}}{(n_X/n_{\rm H})_\odot},
\end{equation}
using the solar abundance table of \citet{1989GeCoA..53..197A}. The simulation tracks H, He, C, N, O, Ne, Mg, Si, and Fe. By definition, $\mathcal{A}_{\rm H}=1$; helium and the tracked metals use their individual cell abundances. For the untracked elements S, Ar, Ca, and Ti, we set $\mathcal{A}_X=\mathcal{A}_{\rm Si}$. Other untracked metals included in the model are assigned $\mathcal{A}_X=\mathcal{A}_{\rm Fe}$. These abundance assignments are part of the emission model and need not reproduce each cell's untracked abundance pattern.

We use 1,000 uniformly spaced energy bins spanning 0.5--2.0 keV. No instrument response is applied. Line broadening is not included in the emissivity calculation, as we focus on band-integrated luminosity rather than spectral line profiles. For energy bin $i$, with lower and upper edges $E_{i,\rm lo}$ and $E_{i,\rm hi}$, \texttt{PyAtomDB} returns the bin-integrated photon emissivity
\begin{equation}\label{eq3}
    \epsilon_i\left(T,\{\mathcal{A}_X\}\right) =
    \int_{E_{i,\rm lo}}^{E_{i,\rm hi}}
    \frac{d\epsilon}{dE}
    \left(T,\{\mathcal{A}_X\}\right)\,dE,
\end{equation}
in $\mathrm{ph\,cm^3\,s^{-1}}$. We convert this quantity to an energy emissivity using
the bin midpoint, $\bar E_i=(E_{i,\rm lo}+E_{i,\rm hi})/2$:
\begin{equation}\label{eq4}
    \Lambda_{\rm gas}\left(T,\{\mathcal{A}_X\}\right)
    \simeq
    C_{\rm keV}
    \sum_{i=1}^{1000}
    \epsilon_i\left(T,\{\mathcal{A}_X\}\right)\bar E_i,
\end{equation}
where the midpoint energies are in keV and $C_{\rm keV}=1.60218\times10^{-9}\, \mathrm{erg\,keV^{-1}}$. Because $\epsilon_i$ is already integrated over its energy bin, no additional bin-width factor appears in the sum.

\begin{figure}[t!]
    \centering
    \includegraphics[width=0.8\columnwidth]{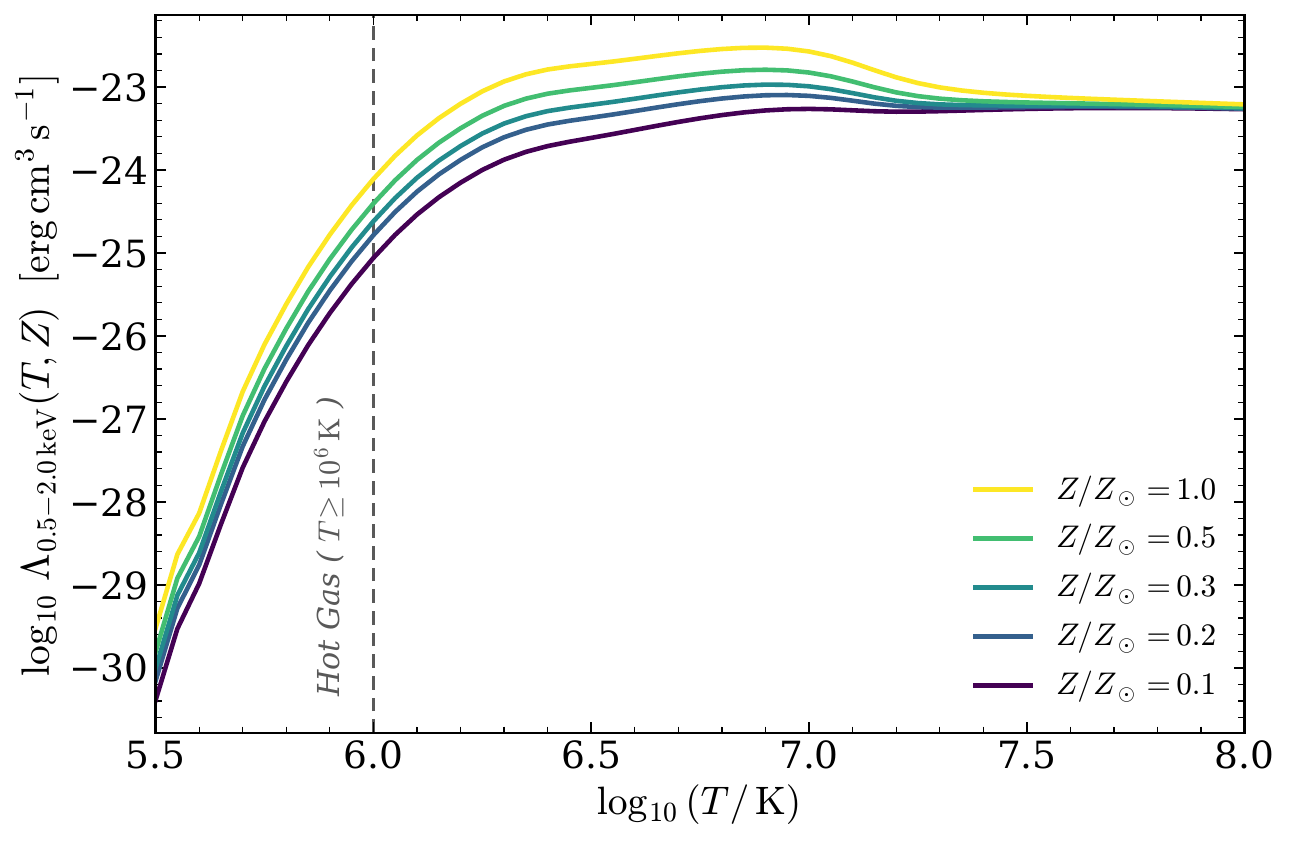}
    \caption{CIE emissivity per emission measure in the 0.5--2.0 keV band, calculated with \texttt{PyAtomDB} \citep{Foster2020Atoms...8...49F}. The illustrative curves adopt the solar abundance pattern of \citet{1989GeCoA..53..197A}. Hydrogen and helium retain their solar abundance ratio, while all metals from Li to Zn are scaled together by $Z/Z_\odot=0.1$, $0.2$, $0.3$, $0.5$, and $1.0$. The cell-by-cell luminosity calculation instead uses the elemental abundances described in the text. The vertical dashed line marks the adopted hot-gas threshold, $T=10^6\,\mathrm{K}$. Metal-line emission produces a strong abundance dependence at the lower temperatures shown. At $\log_{10}(T/\mathrm{K})\gtrsim7.5$, the curves approach one another as the continuum contribution becomes more important.}
\label{fig: coolingfunction}
\end{figure}

Figure~\ref{fig: coolingfunction} illustrates the temperature and abundance dependence of the band emissivity. We evaluate Equation~\ref{eq1} using each selected cell's temperature and elemental abundances and sum the cell luminosities:
\begin{equation}
    L_{\mathrm{X}}
    =
    \sum_{j\in\mathrm{hot\ CGM}}L_{\mathrm{X},j}.
\end{equation}
All reported luminosities are intrinsic, unabsorbed, rest-frame 0.5--2.0 keV luminosities in $\mathrm{erg\,s^{-1}}$.

\section{X-ray Luminosity, Mass, and Star Formation}\label{sec:LX_mass_SFR_relation}

We examine the dependence of $L_{\mathrm{X}}$ on halo mass $M_{500}$, stellar mass $M_\star$, and SFR. We first describe the mass--luminosity relations for the three activity classes and the combined sample. We then test how much of the luminosity--SFR association remains after the specified mass adjustment.

\begin{figure*}[t!]
    \centering
    \includegraphics[width=0.85\textwidth]{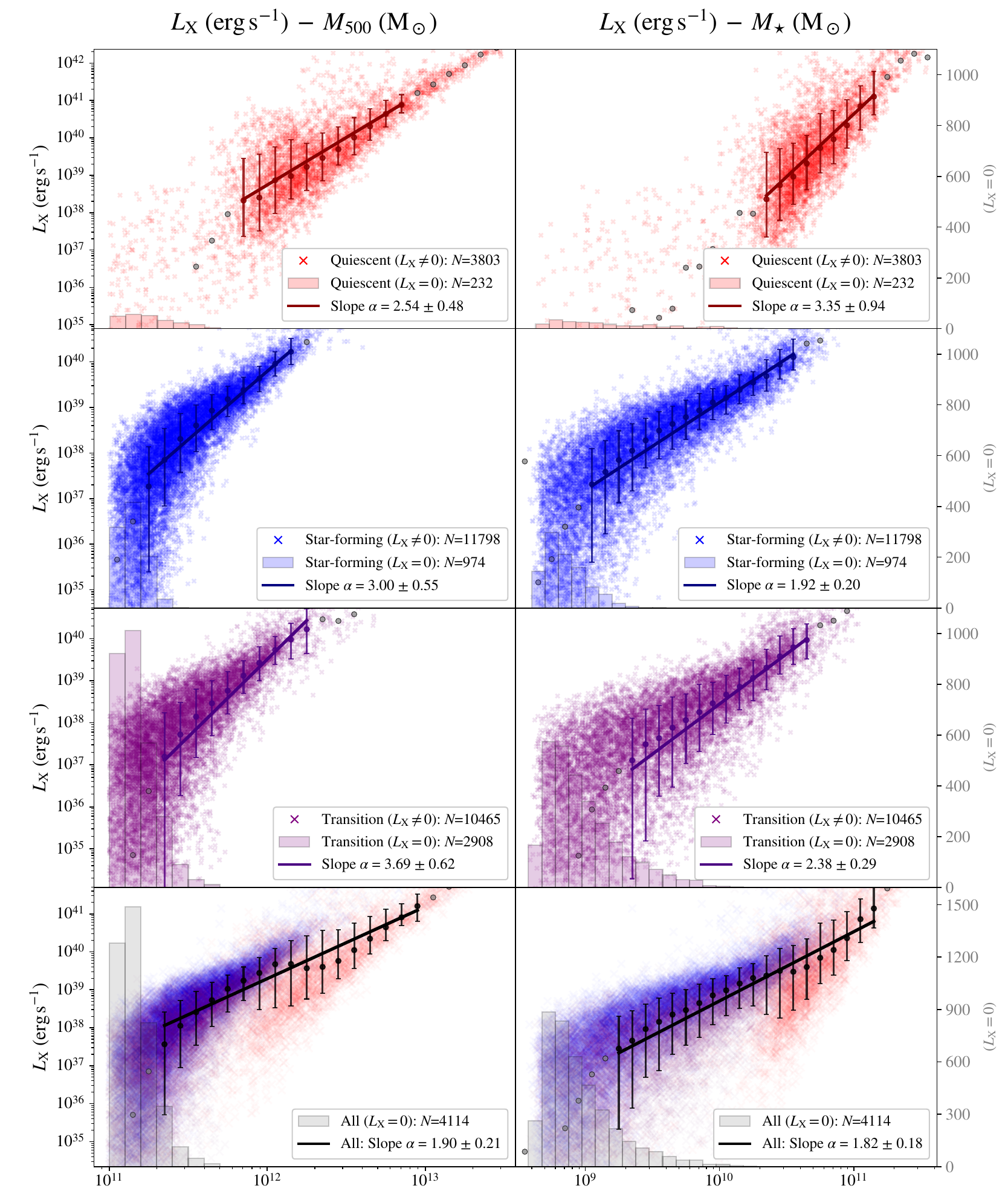}
     \caption{Intrinsic hot-CGM luminosity versus $M_{500}$ (left) and $M_\star$ (right). Rows show quiescent, star-forming, transition, and combined samples, pooling $z=0$, $0.1$, and $0.2$. Crosses show positive-luminosity entries. Histograms show the mass distributions of zero-luminosity entries, with counts on the secondary axes. Points mark linear-space median luminosities, including zeros, at the centers of accepted 0.1~dex mass bins. Bars span the 16th--84th population percentiles, not uncertainty on the median. Accepted bins contain at least 100 entries and have positive 16th-percentile luminosities. Gray points without bars mark positive medians failing either criterion, and nonpositive medians cannot be plotted. Display limits do not exclude entries from the calculation. Solid lines use the median sampled coefficients over the accepted mass ranges. The quoted slope widths are half the 16th--84th percentile spans of their sampled distributions, conditional on the fitting prescription in Section~\ref{subsec:Scaling Relation with Mass: Fitting Method}.}
\label{fig:LX_MassScaling}
\end{figure*}

\subsection{Scaling Relation with Mass: Fitting Method}\label{subsec:Scaling Relation with Mass: Fitting Method}

Figure~\ref{fig:LX_MassScaling} shows the mass--luminosity relations. For each activity class and for the combined sample, we use 0.1 dex bins in $\log_{10}(M/M_\odot)$, where $M$ denotes either $M_{500}$ or $M_\star$. The bin summaries include entries with $L_{\mathrm{X}}=0$. These zeros belong to the defined simulation measurement and are not observational upper limits. The fits summarize the relation over the accepted mass interval, and comparison with other measurements must also account for their aperture, gas selection, and averaging statistic.

For each activity class and for the combined sample, we adopt the linear-space median value of $L_{\mathrm{X}}$ in each 0.1~dex mass bin as the representative value, including entries with $L_{\mathrm{X}}=0$. The bars span the 16th--84th percentiles of the individual luminosities and describe population spread, not uncertainty on the median. We fit bins containing at least 100 entries and having a positive 16th-percentile luminosity, which permits the lower fitting width to be defined in logarithmic units. Bins must also have finite lower and upper fitting widths.

We summarize the accepted bin medians with the relation $y=\alpha x+\beta$, where $x=\log_{10}(M/M_\odot)$ and $y=\log_{10}[L_{\mathrm{X}}/(\mathrm{erg\,s^{-1}})]$. For each bin, $x_k$ is the midpoint of its logarithmic mass interval and $y_k$ is the logarithm of the linear-space median luminosity, including zero values. If $L_{16,k}$, $L_{50,k}$, and $L_{84,k}$ denote the population percentiles of the linear luminosities, the fitting widths are
\begin{equation}
    \sigma_{{\rm lo},k}=\log_{10}(L_{50,k}/L_{16,k}),\qquad
    \sigma_{{\rm hi},k}=\log_{10}(L_{84,k}/L_{50,k}).
\end{equation}
Any nonpositive width is replaced by $10^{-4}$ dex for numerical stability. These widths describe the spread of individual luminosities within a bin, not the sampling uncertainty of its median. We sample the coefficients with \texttt{emcee} \citep{Foreman-Mackey2013PASP..125..306F}, using the affine-invariant ensemble sampler \citep{Goodman2010CAMCS...5...65G}. Our fitting prescription is
\begin{equation}
    \ln\mathcal{L}=-\frac{1}{2}\sum_k \left[\frac{(y_k-\hat y_k)^2}{\sigma_k^2} +\ln(\sigma_k^2)\right],
\end{equation}
where $\hat y_k=\alpha x_k+\beta$ and $\sigma_k=\sigma_{{\rm hi},k}$ for $y_k\geq\hat y_k$, and $\sigma_k=\sigma_{{\rm lo},k}$ otherwise. We adopt uniform priors $-5<\alpha<10$ and $-200<\beta<200$. We initialize 64 ensemble walkers near a weighted-least-squares solution and run each for 4000 steps, discard the first 500 steps, and retain every fifteenth subsequent sample.

We report the median of each sampled coefficient distribution and a symmetrized width equal to half the difference between its 84th and 16th percentiles. These widths are conditional on the adopted fitting prescription and population-based weights. They have not been calibrated as sampling uncertainties of the median relation, and the fit does not account for repeated appearances of the same galaxy lineage across snapshots. We use the fits as descriptive summaries and do not use these widths to assign a significance to differences among the activity classes.

\subsection{Scaling Relation with Mass: Results}\label{subsec:scaling_mass_results}

\begin{table*}[!t]
\centering
\caption{Descriptive Fits to the Median $L_{\mathrm{X}}$--Mass Relations \label{tab:scaling_median}}
\small 
\setlength{\tabcolsep}{6pt} 
\begin{tabular*}{\textwidth}{@{\extracolsep{\fill}} lccccc @{}}
\hline\hline
Population & Mass & Slope & Intercept & $\log_{10}(M/M_\odot)$ & $N_{\rm fit}$ \\
\hline
\textbf{Quiescent} & $M_{500}$ & $2.54 \pm 0.48$ & $8.28 \pm 6.01$ & $[11.85, 12.85]$ & 2991 \\
 & $M_\star$ & $3.35 \pm 0.94$ & $3.73 \pm 10.14$ & $[10.35, 11.15]$ & 3117 \\
\hline
\textbf{Star-forming} & $M_{500}$ & $3.00 \pm 0.55$ & $3.78 \pm 6.53$ & $[11.25, 12.15]$ & 8882 \\
 & $M_\star$ & $1.92 \pm 0.20$ & $19.87 \pm 1.94$ & $[9.05, 10.55]$ & 8949 \\
\hline
\textbf{Transition} & $M_{500}$ & $3.69 \pm 0.62$ & $-4.79 \pm 7.33$ & $[11.35, 12.25]$ & 6123 \\
 & $M_\star$ & $2.38 \pm 0.29$ & $14.60 \pm 2.93$ & $[9.35, 10.65]$ & 6324 \\
\hline
\textbf{All Galaxies} & $M_{500}$ & $1.90 \pm 0.21$ & $16.52 \pm 2.58$ & $[11.35, 12.95]$ & 17025 \\
 & $M_\star$ & $1.82 \pm 0.18$ & $20.46 \pm 1.80$ & $[9.25, 11.15]$ & 18194 \\
\hline
\end{tabular*}
\vspace{1ex}
\parbox{\textwidth}{\footnotesize \textbf{Note.} Fits summarize linear-space median luminosities, including zeros, in accepted 0.1 dex mass bins, using the prescription in Section~\ref{subsec:Scaling Relation with Mass: Fitting Method}. Coefficients are medians of their sampled distributions. The quoted $\pm$ values are symmetrized half-widths, $(q_{84}-q_{16})/2$, conditional on the population-based fitting widths and selected bins. They are not calibrated sampling uncertainties and are not used to assess the significance of differences between activity classes. The log-mass limits are the centers of the lowest- and highest-mass accepted bins. $N_{\rm fit}$ counts positive-luminosity halo--snapshot entries between these centers, inclusive; it is not the number of entries in the accepted bins. The zero values are simulation measurements, not observational nondetections.}
\end{table*}

Table~\ref{tab:scaling_median} summarizes the fitted coefficients and mass ranges for each activity class and the combined sample. All classes show increasing $L_{\mathrm{X}}$ with mass over their fitted intervals. For the halo-mass relation, the transition class has the largest nominal slope, $\alpha=3.69$, followed by the star-forming and quiescent classes, with $\alpha=3.00$ and $2.54$, respectively. The combined sample gives $\alpha=1.90$.

The ordering differs for stellar mass. The quiescent class has the largest nominal slope,
$\alpha=3.35$, followed by the transition and star-forming classes, with $\alpha=2.38$ and $1.92$, respectively. The combined sample gives $\alpha=1.82$. For both mass proxies, the combined-sample slope is smaller than each of the class-specific estimates.

The fitted slopes give the nominal ordering over the class-dependent mass intervals listed in Table~\ref{tab:scaling_median}. The interpretation of their coefficient widths is given in Section~\ref{subsec:Scaling Relation with Mass: Fitting Method}. Appendix~\ref{appendix: A} places the combined-sample relations in the context of earlier observational and simulation measurements.

\begin{table*}[!t]
\centering
\caption{$L_{\mathrm{X}}$--SFR Correlations by Redshift \label{tab:lx_sfr_corr}}
\begin{tabular*}{\textwidth}{@{\extracolsep{\fill}} llccccccc}
\hline\hline
Population & $z$ & $N$ & $r$ & $\rho$ & Partial $r\mid M_{500}$ & Partial $r\mid M_\star$ & Partial $\rho\mid M_{500}$ & Partial $\rho\mid M_\star$ \\
\hline
\multirow{3}{*}{\textbf{Star-forming}} & 0.0 & 3,610 & $+0.81$ & $+0.86$ & $+0.43$ & $+0.31$ & $+0.50$ & $+0.37$ \\
 & 0.1 & 3,981 & $+0.83$ & $+0.88$ & $+0.44$ & $+0.33$ & $+0.51$ & $+0.39$ \\
 & 0.2 & 4,207 & $+0.84$ & $+0.89$ & $+0.47$ & $+0.33$ & $+0.56$ & $+0.40$ \\
\hline
\multirow{3}{*}{\textbf{Transition}} & 0.0 & 3,132 & $+0.72$ & $+0.75$ & $+0.32$ & $+0.28$ & $+0.35$ & $+0.32$ \\
 & 0.1 & 3,500 & $+0.74$ & $+0.78$ & $+0.36$ & $+0.30$ & $+0.42$ & $+0.36$ \\
 & 0.2 & 3,833 & $+0.77$ & $+0.82$ & $+0.37$ & $+0.26$ & $+0.43$ & $+0.33$ \\
\hline
\multirow{3}{*}{\textbf{Quiescent}} & 0.0 & 298 & $+0.37$ & $+0.33$ & $+0.01^{*}$ & $-0.08^{*}$ & $-0.04^{*}$ & $-0.05^{*}$ \\
 & 0.1 & 279 & $+0.40$ & $+0.37$ & $+0.13$ & $+0.07^{*}$ & $+0.16$ & $+0.13$ \\
 & 0.2 & 258 & $+0.36$ & $+0.33$ & $+0.17$ & $+0.11^{*}$ & $+0.16$ & $+0.17$ \\
\hline
\end{tabular*}
\vspace{1ex}
\parbox{\textwidth}{\footnotesize \textbf{Note.} Correlation between X-ray luminosity and instantaneous SFR, evaluated separately for each activity class at $z=0$, $0.1$, and $0.2$. $N$ is the number of entries with both
$L_{\mathrm{X}}>0$ and $\mathrm{SFR}>0$. The coefficients describe these positive-value subsets, not the full populations. $r$ and $\rho$ denote the unadjusted Pearson and Spearman coefficients between logarithmic luminosity and SFR. Partial coefficients adjust for either logarithmic
halo mass or logarithmic stellar mass, as indicated. For a partial Pearson coefficient $r_{\rm p}$, we calculate a two-sided $p$-value using $t=r_{\rm p}\sqrt{(N-3)/(1-r_{\rm p}^{2})}$ and Student's $t$ distribution with $N-3$ degrees of freedom. For partial Spearman coefficients, we apply the same calculation to the ranked variables as an approximation. An asterisk marks a partial coefficient with $p\geq0.05$. The tests assume independent entries within each snapshot, and the estimates from different snapshots are not independent replications. Variation among snapshot estimates is not a confidence interval on a common coefficient.}
\end{table*}

\subsection{\texorpdfstring{Relation between $L_{\mathrm{X}}$ and Star Formation Rate}{Relation between LX and Star Formation Rate}}
\label{subsec: Relation between the L_X and Star Formation Rate}

Motivated by the different distributions of $L_{\mathrm{X}}$ across the activity classes, we examine its relation to the instantaneous SFR measured within 30 physical kpc. We calculate Pearson ($r$) and Spearman ($\rho$) correlations between logarithmic luminosity and logarithmic SFR separately for each class at $z=0$, $0.1$, and $0.2$. The analysis requires both $L_{\mathrm{X}}>0$ and $\mathrm{SFR}>0$. These are positive-value selections within the simulation, not observational detections.

Calculating the coefficients separately at each snapshot avoids treating repeated appearances of the same galaxies as independent entries in a pooled estimate. The estimates from different snapshots nevertheless remain correlated because their samples share galaxy lineages.

We use partial correlations to describe the residual association after adjusting for one mass proxy at a time. For the logarithmic variables and $M=M_{500}$ or $M_\star$,
\begin{equation}\label{eq:partial_corr}
    r_{L_\mathrm{X},\mathrm{SFR}\mid M}
    =
    \frac{r_{L_\mathrm{X},\mathrm{SFR}}-r_{L_\mathrm{X},M}r_{\mathrm{SFR},M}}{\sqrt{(1-r_{L_\mathrm{X},M}^2)(1-r_{\mathrm{SFR},M}^2)}}.
\end{equation}
Here $r_{a,b}$ is the Pearson coefficient between the specified logarithmic variables. We calculate the partial Spearman coefficient using
the same expression after replacing the variables by their ranks, assigning average ranks to tied values.

The Pearson statistic adjusts for a linear dependence on log mass, while the rank-based statistic adjusts for a linear dependence on ranked mass.
Neither removes every possible mass dependence or establishes a causal effect of star formation on the CGM.

\begin{figure}[t!]
    \centering
    \includegraphics[width=0.75\columnwidth]{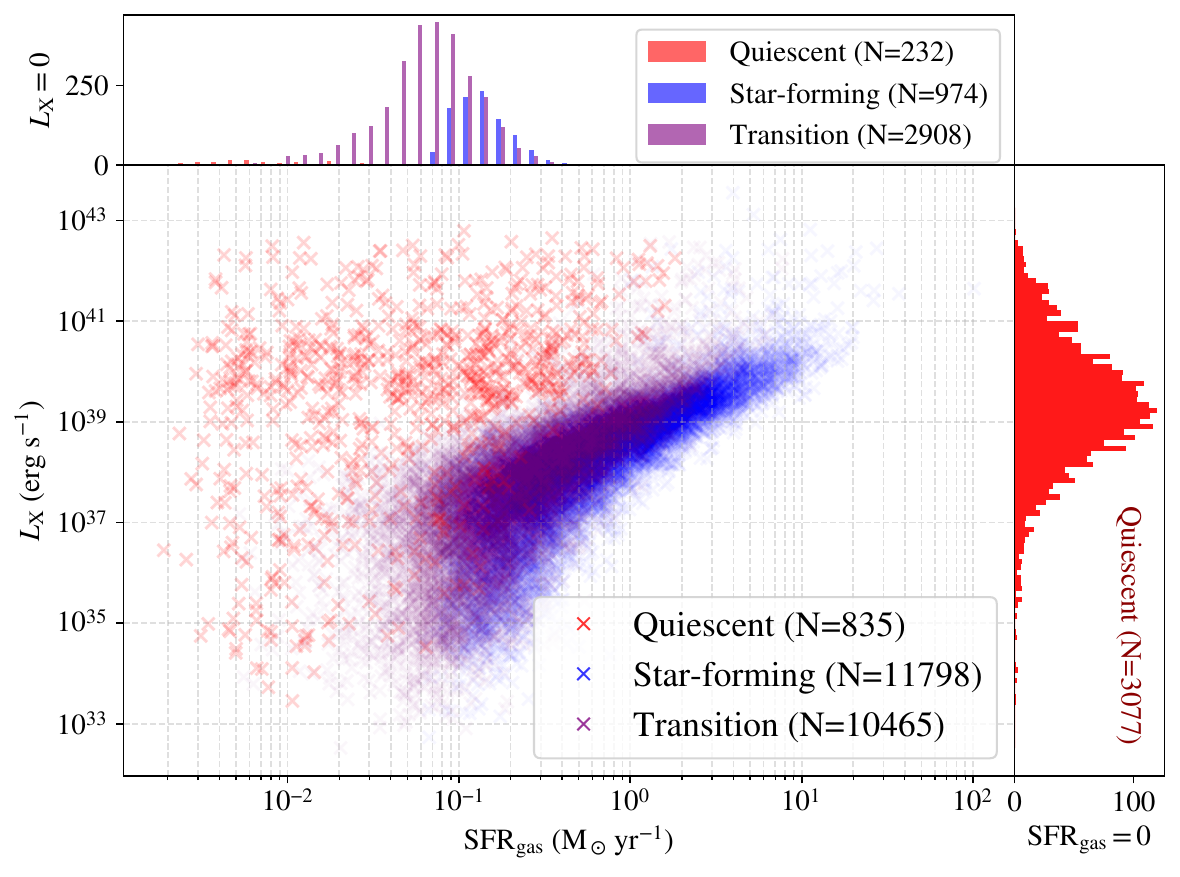}
    \caption{Intrinsic hot-CGM X-ray luminosity versus instantaneous SFR measured within 30 physical kpc for the quiescent, star-forming, and transition samples. The figure pools halo--snapshot entries at $z=0$, $0.1$, and $0.2$. The main panel shows entries with both $L_{\mathrm{X}}>0$ and $\mathrm{SFR}>0$. The top histogram shows positive SFRs of entries with $L_{\mathrm{X}}=0$, and the right histogram shows positive luminosities of entries with $\mathrm{SFR}=0$. Of the 4035 quiescent entries, 835 have both quantities positive and 109 have both equal to zero. The latter cannot be displayed on the logarithmic axes but are included in the quoted counts for both marginal histograms. These two marginal counts therefore overlap and should not be added as disjoint samples. The unadjusted association is positive for all three classes but weaker for the quiescent subset. Per-snapshot correlation coefficients are reported in Table~\ref{tab:lx_sfr_corr}.}
    \label{fig:LX_SFRScaling}
\end{figure}

Figure~\ref{fig:LX_SFRScaling} shows the pooled luminosity--SFR distribution and the zero-valued entries excluded from its main panel. Only 835 of the 4,035 quiescent halo--snapshot entries have both positive luminosity and positive instantaneous SFR. The quiescent correlation therefore describes the subset with residual star formation, not the full quiescent population. The star-forming and transition correlations are also restricted to positive-luminosity entries. Within these selected samples, the quiescent distribution is more scattered and its raw positive association is weaker than those of the other two classes.

Table~\ref{tab:lx_sfr_corr} reports the coefficients separately for the three snapshots. The unadjusted Pearson coefficient is $r=0.81$--$0.84$ for the star-forming class, $0.72$--$0.77$ for the transition class, and
$0.36$--$0.40$ for the positive-luminosity, positive-SFR quiescent subset. The Spearman coefficients show the same ordering.

These ranges describe variation among snapshot estimates, not confidence intervals on a common coefficient. We interpret the ordering descriptively, since differences among the classes can also reflect their
selected mass and SFR ranges.

After adjustment for halo mass, the star-forming Pearson coefficient decreases from $0.81$--$0.84$ to $0.43$--$0.47$ across the three snapshots.
A positive residual association also remains in the transition sample, with corresponding coefficients of $0.32$--$0.37$.
For the quiescent subset, both Pearson and Spearman partial coefficients are small, with magnitudes below approximately $0.18$ across the reported mass controls and snapshots. Their values vary with the chosen mass proxy and correlation statistic.

These small quiescent coefficients do not establish that the residual association is absent. They apply to the subset with positive instantaneous
SFR and positive X-ray luminosity, rather than to the full quiescent population.
Their sensitivity to the SFR averaging timescale has not been established for the same sample and mass adjustments. More generally, the residual associations do not isolate an independent or causal effect of star formation on CGM emission.

\section{Radial Structure of the CGM in Milky Way scale Halos}\label{sec:Star-Forming vs Quiescent galaxy in Milky Way mass galaxy}

We next compare the radial gas and emission profiles of star-forming and quiescent galaxies in a restricted halo-mass interval, $10^{11.75}<M_{500}/M_\odot<10^{12.25}$, and refer to these systems as Milky Way-scale halos. This interval brackets observational estimates of the Milky Way, $M_{200}\approx1.1\times10^{12}\,M_\odot$ \citep{Bland-Hawthorn2016ARA&A..54..529B}, which corresponds to $\log_{10}(M_{500}/M_\odot)\approx11.9$ for a Navarro--Frenk--White profile with a standard concentration--mass relation \citep{Hu2003ApJ...584..702H}. The selection limits the mass range but does not match the halo-mass distributions of the two activity classes.

The sample contains 653 quiescent and 607 star-forming galaxies at $z=0$, 523 and 737 at
$z=0.1$, and 453 and 703 at $z=0.2$, respectively. We pool the three snapshots for the
profiles below, giving 1,629 quiescent and 2,047 star-forming halo--snapshot entries.

We divide the spherical CGM region, $0.15R_{500}<r\leq R_{500}$, into shells of width $0.05R_{500}$. For each halo, we calculate the X-ray luminosity and mass-weighted gas properties in these shells. We use the same hot-gas selection, $T\geq10^6\,\mathrm{K}$, for the X-ray luminosities, hot-gas masses, thermodynamic profiles, and hot-gas shell occupancy.

We also calculate the cumulative electron-pressure signal,
\begin{equation}
    S_{\mathrm{tSZ}}(<r)
    =
    \frac{\sigma_{\mathrm{T}}}{m_{\mathrm{e}}c^2}
    \sum_{0.15R_{500}<r_i\leq r}
    n_{\mathrm{e},i}k_{\mathrm{B}}T_iV_i,
\end{equation}
where $V_i$ is the physical volume of gas cell $i$. This quantity has units of physical area and includes all selected gas cells regardless of temperature. Following the fixed-distance convention of \citet{Junhan2022ApJ...926..179K}, we report
\begin{equation}
    \widetilde{Y}_{\mathrm{tSZ}}(<r)
    =
    E(z)^{-2/3}
    \frac{S_{\mathrm{tSZ}}(<r)}
    {(500\,\mathrm{Mpc})^2},
\end{equation}
converted from radians squared to arcmin$^2$. The reference distance is a normalization convention, not the angular-diameter distance of an object at $z=0$.

\begin{figure*}[t!]
    \centering
    \includegraphics[width=\textwidth]{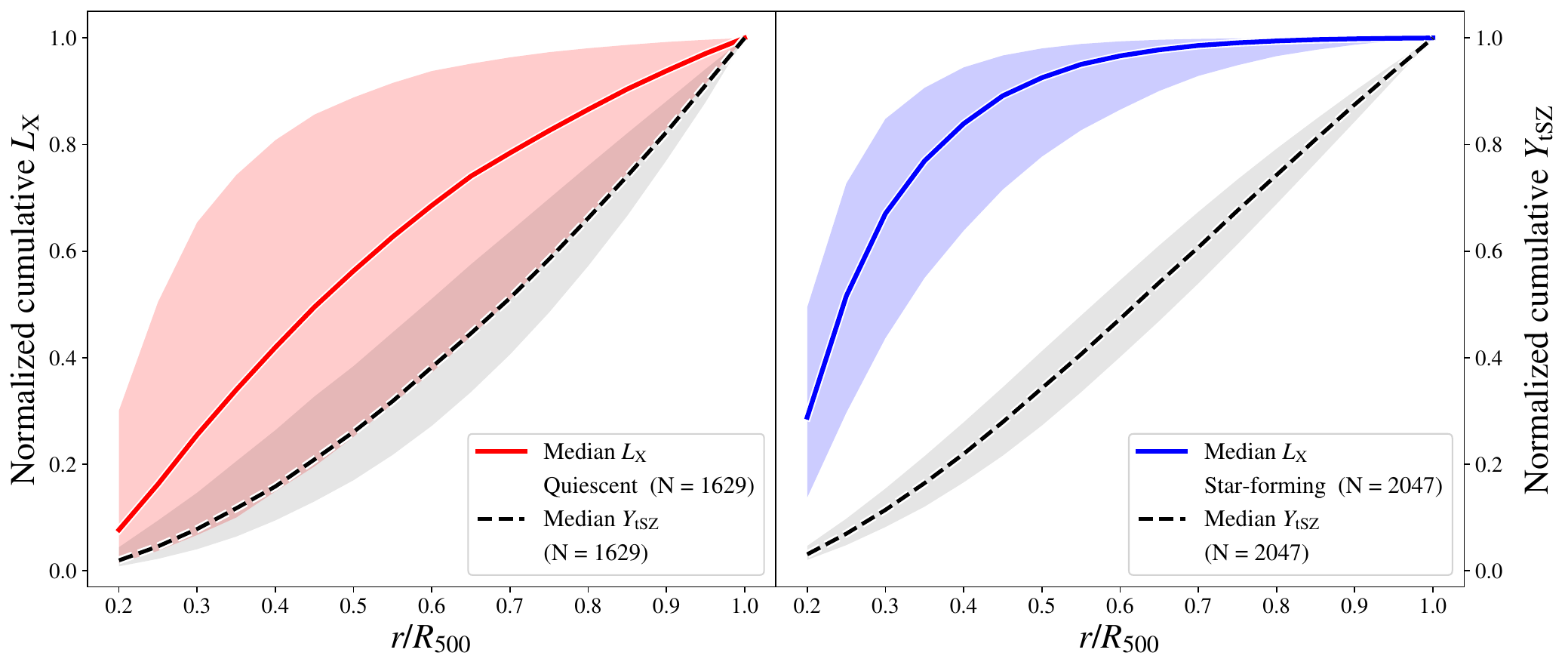}
    \caption{Normalized cumulative profiles of intrinsic X-ray luminosity, $L_{\mathrm{X}}$, and integrated tSZ signal, $\widetilde{Y}_{\mathrm{tSZ}}$, in the spherical shell $0.15R_{500}<r\leq R_{500}$. The left and right panels show the quiescent and star-forming samples, respectively. The profiles pool 1629 quiescent and 2047 star-forming halo--snapshot entries at $z=0$, $0.1$, and $0.2$. Some galaxies contribute at more than one snapshot. Each cumulative profile is divided by its own signal in the full shell before the population median is calculated. Colored solid lines show the median X-ray profiles, and black dashed lines show the median tSZ profiles. Shading spans the 16th--84th percentiles of individual normalized profiles, not uncertainty on the median. Normalization requires a finite, positive total signal, with this condition applied separately to the two observables. All entries in the stated samples satisfy both conditions. The X-ray calculation includes gas with $T\geq10^6\,\mathrm{K}$, whereas the tSZ calculation includes all selected gas cells regardless of temperature. X-ray emission is more centrally concentrated within the analyzed shell, particularly in the star-forming sample. The samples occupy the same halo-mass interval but are not matched in their halo-mass distributions.}
    \label{fig:LX_RadialProfile}
\end{figure*}

We calculate the cumulative $L_{\mathrm{X}}$ and $\widetilde{Y}_{\mathrm{tSZ}}$ up to each radius and normalize each profile to its own value at $R_{500}$. We then calculate the median and 16th--84th percentile range across halo--snapshot entries. Figure~\ref{fig:LX_RadialProfile} shows the resulting profiles for the quiescent and star-forming samples. Normalization separates profile shape from amplitude. The shading shows the spread among individual halo–snapshot profiles.

We obtain the hot and all-phase gas masses by summing cell masses in the same core-excised shell, with and without the hot-gas temperature cut, respectively. Thermodynamic profiles are conditional on a shell containing hot gas. We report the corresponding shell occupancy separately.

\begin{figure*}[t!]
    \centering
    \includegraphics[width=\textwidth]{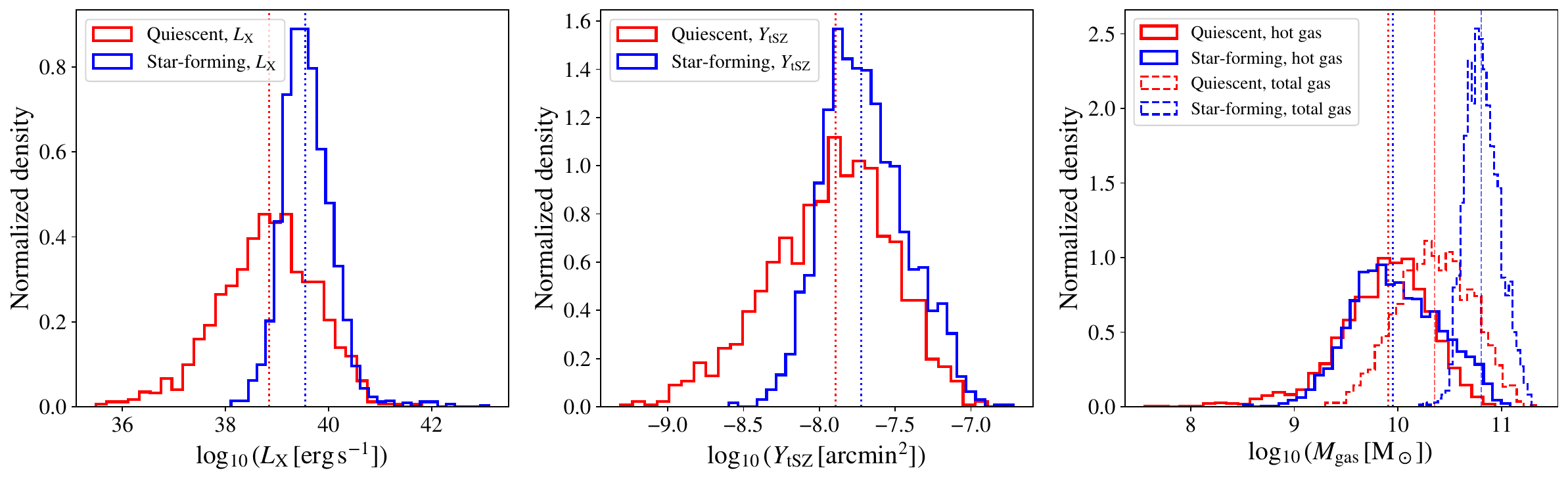}
    \caption{Distributions of intrinsic X-ray luminosity (left), integrated tSZ signal (middle), and gas mass (right) in the spherical shell $0.15R_{500}<r\leq R_{500}$ for the Milky Way scale sample. The red and blue curves represent quiescent and star-forming objects, respectively, pooled over $z=0$, $0.1$, and $0.2$. The parent samples contain 1629 quiescent and 2047 star-forming halo--snapshot entries. The quantities are shown in logarithmic units, and the vertical dotted lines mark their medians. The tSZ signal uses the fixed-distance normalization defined in the text. In the right panel, solid curves show the mass of gas with $T\geq10^6\,\mathrm{K}$ and dashed curves show the all-phase CGM gas mass. The star-forming sample has a higher median X-ray luminosity and all-phase gas mass, while the median hot-gas masses are similar. These are comparisons within a common mass interval, not between samples matched in halo mass.}
    \label{fig:LX_Ytsz_GasMass_histogram}
\end{figure*}

Figure~\ref{fig:LX_RadialProfile} compares the normalized cumulative profiles. X-ray emission is more concentrated toward the inner edge of the analyzed shell than the tSZ signal in both populations.
The median cumulative X-ray profile reaches 90\% of the shell luminosity at approximately $0.85R_{500}$ for quiescent galaxies and $0.46R_{500}$ for star-forming galaxies. These radii are read from the median cumulative curves, rather than calculated as medians of individual 90\%-enclosure radii. They describe emission outside $0.15R_{500}$ and do not include the excised central region.

The normalized tSZ profiles rise more gradually and have similar shapes in the two samples. The quiescent X-ray profiles show greater variation
in radial shape. This comparison describes the selected populations
without removing differences in their halo-mass distributions.

Figure~\ref{fig:LX_Ytsz_GasMass_histogram} shows the corresponding absolute distributions. The medians of $\log_{10}[L_{\mathrm{X}}/(\mathrm{erg\,s^{-1}})]$ are 38.85 and 39.55 for the quiescent and star-forming samples, respectively. The quiescent median is $0.70$~dex (a factor of about five) below the star-forming median, and the quiescent distribution is roughly twice as broad. The 16th--84th percentile half-widths are $0.94$ and $0.45$~dex, respectively. The median redshift-scaled tSZ signal, $\widetilde{Y}_{\mathrm{tSZ}}$, is lower in the quiescent sample by 0.17 dex, or a factor of approximately 1.5. Its contrast is therefore smaller than the luminosity contrast, although the distributions are not identical.

The median values of $\log_{10}(M_{\mathrm{hot}}/M_\odot)$ are 9.90 and 9.95 for the quiescent and star-forming samples, with 16th--84th percentile half-widths of 0.42 and 0.45 dex, respectively.
The all-phase gas masses in the same shell differ more, with median values of $\log_{10}(M_{\mathrm{gas}}/M_\odot)$ of 10.36 and 10.81. Thus the luminosity contrast is not accompanied by a comparable contrast in the median mass above the hot-gas threshold.

These summaries describe samples that are not matched in halo mass. The quoted widths measure variation within each population, not uncertainty on the difference between their medians.

The mass-weighted thermodynamic profiles in Appendix~\ref{appendix: B} provide further context for the luminosity difference. These profiles include only shells containing gas with $T\geq10^6\,\mathrm{K}$. The star-forming sample has higher hot-gas density, with the largest contrast in the inner shells. The temperature profiles do not show a simple offset: the median hot-gas temperature of star-forming galaxies declines with radius, from $T\approx2.1\times10^6$~K in the inner CGM to $T\approx1.2\times10^6$~K near $R_{500}$. Quiescent galaxies have a nearly flat profile around $T\approx(1.8$--$1.9)\times10^6$~K. The temperature ordering therefore changes with radius, and neither population is uniformly hotter throughout the shell.

To show how the selection entering these thermodynamic profiles changes with radius, we define shell occupancy as the fraction of halo--snapshot entries containing at least one gas cell with $T\geq10^6\,\mathrm{K}$ in
a given shell. Both populations have occupancies above 97\% in the
inner CGM, and their occupancies are similar around $0.6$--$0.65R_{500}$.
The difference grows farther out. In the outermost shell, $0.95R_{500}<r\leq R_{500}$, the occupancy is 98.8\% for quiescent galaxies and 84.5\% for star-forming galaxies. Considering the shells whose outer edges are at or beyond $0.5R_{500}$, 18.0\% of the star-forming sample and 3.7\% of the quiescent sample have at least one shell containing no hot gas cells.

The outer thermodynamic profiles therefore describe a smaller fraction of the star-forming sample. An empty hot-gas shell means that no selected cell
exceeds the adopted temperature threshold. It does not show that all gas is absent or has been expelled.
Occupancy also depends on cell sampling and shell width. We use it to describe the selection entering the profiles, not as a resolution-independent measure of the extent of a hot atmosphere. These measurements end at $R_{500}$ and do not constrain the gas distribution beyond that radius.

\section{DISCUSSION}\label{sec:discussion}

\subsection{\texorpdfstring{Scaling Relation between $L_{\mathrm{X}}$}{LX} and Mass }\label{subsec:disc_scaling}

The rank ordering of the fitted slopes differs between the $L_{\mathrm{X}}$--$M_{500}$ and $L_{\mathrm{X}}$--$M_\star$ relations. Among the three activity classes, quiescent galaxies have the shallowest fitted relation with halo mass and the steepest with stellar mass. This ordering refers to the point estimates over the different accepted fitting intervals.

The mapping between stellar and halo mass may contribute to this behavior. Activity-dependent stellar-to-halo mass relations have been reported observationally \citep[e.g.,][]{Mandelbaum2016MNRAS.457.3200M,Tinker2013ApJ...778...93T,Rodriguez-Puebla2015ApJ...799..130R}. Such a difference could affect the relation with stellar mass, but neither a normalization offset nor a narrower fitting interval by itself explains the slope ordering. Establishing that explanation would require examining the mass dependence of the mapping and comparing the relations over a common mass interval.

The fitted relations for the individual activity classes are steeper than the combined-sample relation. Population mixing is one possible contributor: the class fractions and luminosity distributions vary with halo mass, so the median relation of the combined sample need not have the same slope as the relations within its components. The three activity classes overlap in halo mass. Differences in accepted fitting range and normalization must therefore be considered alongside any intrinsic class dependence. A comparison over the common mass interval would help separate these effects.

The combined-sample fit provides a compact description of the intrinsic luminosity--mass relation over the accepted mass range. It is not, by itself, a test of a universal self-similar slope. In the soft X-ray band, the expected mass dependence changes with temperature and abundance, and the fixed $10^6$~K threshold further changes the emitting gas fraction with halo mass. We therefore do not assign a formal significance to a difference from a single reference slope or infer an amount of feedback heating from that difference.

\subsection{\texorpdfstring{$L_{\mathrm{X}}$}{LX} and Star Formation Rate}

Among objects with both $L_{\mathrm{X}}>0$ and $\mathrm{SFR}>0$, the raw luminosity--SFR correlation is stronger in the star-forming and transition samples than in the quiescent sample. A positive residual association remains in the first two classes after the specified linear adjustment for halo mass, although its amplitude is substantially smaller than the raw correlation. This is a mass-adjusted association within a selected sample, not a demonstration of independence from every possible mass dependence.

For quiescent galaxies with positive X-ray luminosity and instantaneous SFR, the residual coefficients are small and vary with the chosen mass proxy and correlation statistic. Most quiescent galaxies fall outside this subset because at least one quantity is zero. Comparisons across activity classes must also account for their different sampled SFR ranges. Sensitivity to the SFR averaging timescale remains untested for the same sample and mass adjustments.

Stellar winds and supernova-driven outflows offer a possible connection between recent star formation and the inner hot CGM. Variations in gas supply could also affect both SFR and X-ray emission. Distinguishing these possibilities, and assessing the contributions of black-hole feedback and accretion heating, requires direct measurements of energy input and gas flows beyond the correlations examined here.

\subsection{Physical Implications of X-ray Emission in Milky Way Mass Systems}

Within the selected Milky Way scale interval, star-forming galaxies have brighter and more centrally concentrated X-ray emission than quiescent galaxies. Their median luminosities differ by 0.70~dex, whereas their median hot-gas masses differ by only 0.05~dex. In contrast to the X-ray profiles, the normalized tSZ profiles have similar shapes, and the median redshift-scaled tSZ signals differ by 0.17~dex. The tSZ measurement integrates electron pressure from all selected gas in the core-excised shell, with the normalization defined in Section~\ref{sec:Star-Forming vs Quiescent galaxy in Milky Way mass galaxy}. The two activity samples occupy the same selected mass interval but retain different halo-mass distributions.

The star-forming sample has higher median hot-gas densities in the inner CGM, consistent with its more centrally concentrated X-ray emission. At the cell level, the luminosity is $L_{\mathrm{X}}=\sum_i n_{\mathrm{e},i}n_{\mathrm{H},i}\Lambda(T_i,{\mathcal{A}_{X,i}})V_i$, so temperature and elemental abundances also contribute through the emissivity. The soft-band emissivity is particularly sensitive to temperature near the adopted threshold. Quantifying the relative contributions requires a cell-level emissivity decomposition beyond the mass-weighted profiles presented here.

Feedback and variations in gas supply could both produce differences in the inner gas reservoir. Previous TNG studies have linked reduced central gas content to the low-accretion kinetic mode of black-hole feedback \citep{Truong2020MNRAS.494..549T}. Across seven simulation suites, \citet{2025ApJ...993..125S} find X-ray-bright outflows in models with AGN feedback and in some models with stellar feedback alone. Their analysis demonstrates the value of combining X-ray morphology with temperature and velocity diagnostics. Testing the feedback interpretation for our samples would require corresponding measurements of energy injection and gas transport.

Overall, X-ray luminosity provides an emissivity-weighted view of the hot CGM, while the tSZ signal measures the pressure contribution of all selected gas. Combining these measurements with hot-gas masses offers complementary information on CGM structure \citep{Battaglia2019BAAS...51c.297B}. Observational comparisons would require matching the projected aperture, galaxy selection, halo-mass distribution, and averaging procedure.

\subsection{Comparison with Previous Studies}

Earlier TNG studies have already demonstrated a dependence of X-ray luminosity on star-formation activity \citep{Oppenheimer2020ApJ...893L..24O, Truong2020MNRAS.494..549T}. Comparisons with our results require accounting for differences in gas selection, aperture, and activity definition. \citet{Oppenheimer2020ApJ...893L..24O} select the upper and lower thirds of the sSFR distribution within their stellar-mass bins. Although this differs from our fixed quiescent threshold, their low-sSFR TNG selections lie below our threshold in both mass bins. A dilution of their TNG quiescent sample by galaxies in our transition class is therefore not established by the percentile definition alone. The EAGLE and TNG selections should be assessed separately.

\citet{Truong2020MNRAS.494..549T} identify quenched galaxies using an offset of at least 1~dex below the SFMS. Our fixed sSFR threshold is a different operational definition, and our star-forming sample is further restricted to galaxies on or above the fitted SFMS. The luminosity contrasts should therefore be compared with these differences in selection in mind.

The scaling relations shown in Appendix~\ref{appendix: A} provide broader context, including the FIRE calculations of \citet{Voort2016MNRAS.463.4533V} and the eROSITA stacks of \citet{Zhang2024A&A...690A.268Z}. These relations use different apertures, mass ranges, sample selections, and summary statistics. Their slope ordering is therefore not evidence for a difference in heating. A quantitative comparison would require recomputing matched quantities, and the present overlay is intended only as a qualitative reference.

The activity-dependent luminosity contrasts agree qualitatively in sign. \citet{Truong2020MNRAS.494..549T} find that star-forming galaxies near $M_\star\approx10^{10.7}M_\odot$ are approximately an order of magnitude brighter than quenched galaxies in their adopted X-ray measurement. \citet{Oppenheimer2020ApJ...893L..24O} report contrasts of approximately 2.5 in EAGLE and 15 in TNG100 for their lower stellar-mass bin. Our Milky Way scale halo selection gives a median contrast of approximately five in the core-excised hot-gas luminosity. These are not measurements of an identical sample or observable. Their common sign is informative, but the different amplitudes cannot be attributed to aperture, resolution, or classification without a matched calculation.

\citet{Truong2020MNRAS.494..549T} connect lower X-ray luminosity to reduced gas content in the central galaxy and to black-hole kinetic feedback. In TNG, the transition between thermal and kinetic feedback depends on accretion state and black-hole mass, and $M_{\mathrm{BH}}\sim10^8M_\odot$ is a characteristic population scale, not a universal hard switch \citep{Weinberger2017MNRAS.465.3291W}. \citet{Oppenheimer2020ApJ...893L..24O} likewise relate the X-ray contrast to differences in CGM gas fraction.

Our all-phase gas mass in the $0.15$--$1.0R_{500}$ shell is lower in the quiescent sample, while the median mass above $10^6$~K is similar between the classes. These measurements concern a different aperture and phase selection from a central total-gas fraction. They do not show that the gas removed in the earlier studies was exclusively cooler gas, or that the hot phase in our sample was redistributed rather than lost and replenished. The contribution here is a joint description of radial X-ray concentration, pressure, and temperature-selected gas content for the selected populations. This comparison separates gas by temperature and radius without reconstructing its history or isolating an effect of activity at fixed halo mass.

Spatially extended TNG X-ray predictions are also available from \citet{2023MNRAS.525.1976T}, who examine metal-line emission and compare the radial density, temperature, and metallicity of gas around star-forming and quiescent galaxies. Our analysis complements that work by relating broadband X-ray concentration to the pressure-weighted profile and the mass of gas above a specified temperature threshold in the same core-excised shell. In EAGLE and SIMBA, \citet{2025ApJ...994...89G} find that the relative roles of density, temperature, and abundance vary with radius. We have not separated those contributions in our TNG sample, so the present profiles cannot establish that the same component dominates at every radius.

Using Magneticum, \citet{2025A&A...695A...2V} find steeper gas-only X-ray surface-brightness profiles for star-forming galaxies among systems fitted by single $\beta$ models, qualitatively resembling our profile-shape ordering. Their Milky Way scale subsamples matched in mean stellar mass nevertheless show no significant difference in the mean gas surface-brightness profiles. Our comparison uses medians of individually normalized cumulative luminosity profiles in a core-excised spherical shell, so the profile statistic and sample selection are important distinctions.

Recent observational comparisons further emphasize the importance of mass selection. The TNG300 forward modeling of \citet{2025A&A...703A.137S} shows why matching stellar mass and redshift does not necessarily match the underlying halo-mass distributions. \citet{Zhang2025A&A...693A.197Z} find similar X-ray luminosities for star-forming and quiescent systems at fixed inferred halo mass above $\log_{10}(M_{200\mathrm{m}}/M_\odot)=12.5$. The different mass definition, projected aperture, and stack statistic prevent a direct comparison with our core-excised $M_{500}$ selection. The sensitivity of inferred CGM gas masses to temperature, abundance, and halo-mass assumptions found by \citet{2026A&A...706A.102Z} also shows why luminosity cannot be converted into gas mass without an emissivity model. At higher masses, \citet{2026arXiv260621495A} provide additional APEC-based scaling and core-excision analyses for groups. Those results supply context at the upper end of our sample, rather than a numerical validation of the present Milky Way scale contrast.

Recent stacks of nearby $L^*$ galaxies show a steeper radial decline of X-ray emission in the higher-SFR subset \citep{2026arXiv260116499H}. This resembles the direction of our profile-shape difference, but their higher-SFR sample is fainter when the emission is integrated over 10--200 kpc. Their optical selection, projected aperture, and stacking statistic differ from our halo-mass selection and intrinsic spherical-shell measurement. The comparison therefore suggests qualitative similarity in radial shape, not agreement in the luminosity ordering.

Using 32 star-forming Milky Way analogues in TNG50, \citet{2026ApJ..1004..184Z} find broadly realistic inner-halo X-ray emission but surface-brightness profiles that fall below eROSITA stacks of SDSS galaxies at large radii. They also report weaker O~VIII absorption than observed in the Galactic halo. Their results illustrate how agreement in integrated luminosity can coexist with discrepancies in radial structure and the temperature distribution.

These results describe the adopted TNG100-2 measurement and have not been tested for numerical convergence across resolutions. The central-subhalo gas-count cut can also omit gas-poor systems. In addition, the fixed $10^6$~K threshold selects only part of the CGM temperature distribution, and the CIE emission model does not include photoionization or non-equilibrium ionization. The reported gas masses apply to the specified phases and aperture, while the luminosities also depend on the plasma model. Resolution and gas selection deserve particular care when interpreting low-mass halos and the radial occurrence of hot gas.

\section{SUMMARY \& CONCLUSIONS}\label{sec:conclusion}

We have calculated the intrinsic 0.5--2~keV luminosity of the hot CGM in TNG100-2 at $z=0$, 0.1, and 0.2. The calculation includes gas with $T\geq10^6$~K in the core-excised spherical shell between $0.15R_{500}$ and $R_{500}$. We compare three activity-selected samples and examine the radial profiles and gas masses of the high-activity and quiescent Milky Way scale subsets. Our main results are as follows.

\begin{enumerate}[leftmargin=*]

\item X-ray luminosity increases with both halo mass and stellar mass. The activity-selected samples occupy different parts of these relations, and their fitted slopes depend on the accepted mass intervals. We therefore interpret the fits as descriptions of the selected populations, rather than as a measurement of a universal soft-band slope or a direct test of feedback heating.

\item Among galaxies with positive $L_{\mathrm{X}}$ and SFR, the star-forming and transition samples retain a positive luminosity--SFR association after a linear adjustment for halo mass. The residual association is weaker in the quiescent subset. These correlations describe the selected positive-value samples. They do not identify the heating source or establish the absence of an SFR dependence in the full quiescent population.

\item Within $10^{11.75}<M_{500}/M_\odot<10^{12.25}$, the star-forming sample has more centrally concentrated X-ray emission than the quiescent sample. In the pooled sample, the median cumulative profiles enclose 90\% of their shell luminosity by approximately $0.46R_{500}$ and $0.85R_{500}$, respectively. The contrast in the pressure-weighted profiles is smaller. The comparison applies to the core-excised shell $0.15R_{500}<r\leq R_{500}$ within the stated halo-mass interval. The two activity samples have not been matched in halo mass.

\item The pooled Milky Way scale samples differ in median X-ray luminosity by about 0.70~dex, while their median masses above $10^6$~K are similar. The quiescent sample has a lower median all-phase gas mass in the same shell. The direction of the luminosity contrast agrees qualitatively with earlier TNG predictions \citep{Truong2020MNRAS.494..549T,Oppenheimer2020ApJ...893L..24O}, but the different apertures and selections prevent a direct comparison of its amplitude. Our result concerns the low-redshift spatial and temperature-phase distributions of the selected populations.

\end{enumerate}

Soft X-ray luminosity is sensitive to both the amount of hot gas and its spatial and thermodynamic structure. In our selected samples, a fivefold luminosity contrast accompanies similar median hot-gas masses and different radial emission profiles. Comparing luminosity and radial concentration in activity-selected galaxies with comparable halo-mass distributions would provide an observational test. The corresponding simulation measurements would need to match the observed apertures, unresolved-source treatment, and averaging procedure.

\begin{acknowledgments}
This work was supported by KAIST through its Undergraduate Research Participation (URP) Program and the associated follow-up research program. This work was supported by the National Research Foundation of Korea (NRF) grants funded by the Korea government (MSIT) (RS-2024-00340759 and RS-2025-16302968). We thank our colleagues in the ObsCAL group and participants in meetings of the Korean Astronomical Society (KAS) and the Korean Physical Society (KPS) for helpful discussions.
\end{acknowledgments}

Disclosure Statement: During the preparation of this manuscript, Claude Code (Anthropic) and ChatGPT (OpenAI) were used to assist with editing the language and organization of the author's original writing, and refining plotting scripts for figure presentation. The authors reviewed and revised the assisted text and verified code modifications and figures against analysis outputs. The authors take full responsibility for the analyses, interpretations, and final content of this paper.

\software{
Astropy \citep{2013A&A...558A..33A, 2018AJ....156..123A, 2022ApJ...935..167A},
illustris\_python (\url{https://github.com/illustristng/illustris_python}),
PyAtomDB \citep{Foster2020Atoms...8...49F},
NumPy \citep{2020Natur.585..357H},
SciPy \citep{2020NaMet..17..261V},
Matplotlib \citep{2007CSE.....9...90H},
emcee \citep{Foreman-Mackey2013PASP..125..306F}
}

Data availability: The TNG100-2 simulation data are available through the IllustrisTNG public data release \citep{2019ComAC...6....2N} (\url{https://www.tng-project.org/data/}). This work uses snapshots 99, 91, and 84, the associated group catalogs, and the SubLink merger trees.

\appendix

\section{Alternative Scaling Fits and Comparison with Published Relations} \label{appendix: A}

\begin{figure*}[t!]
    \centering
    \includegraphics[width=\textwidth]{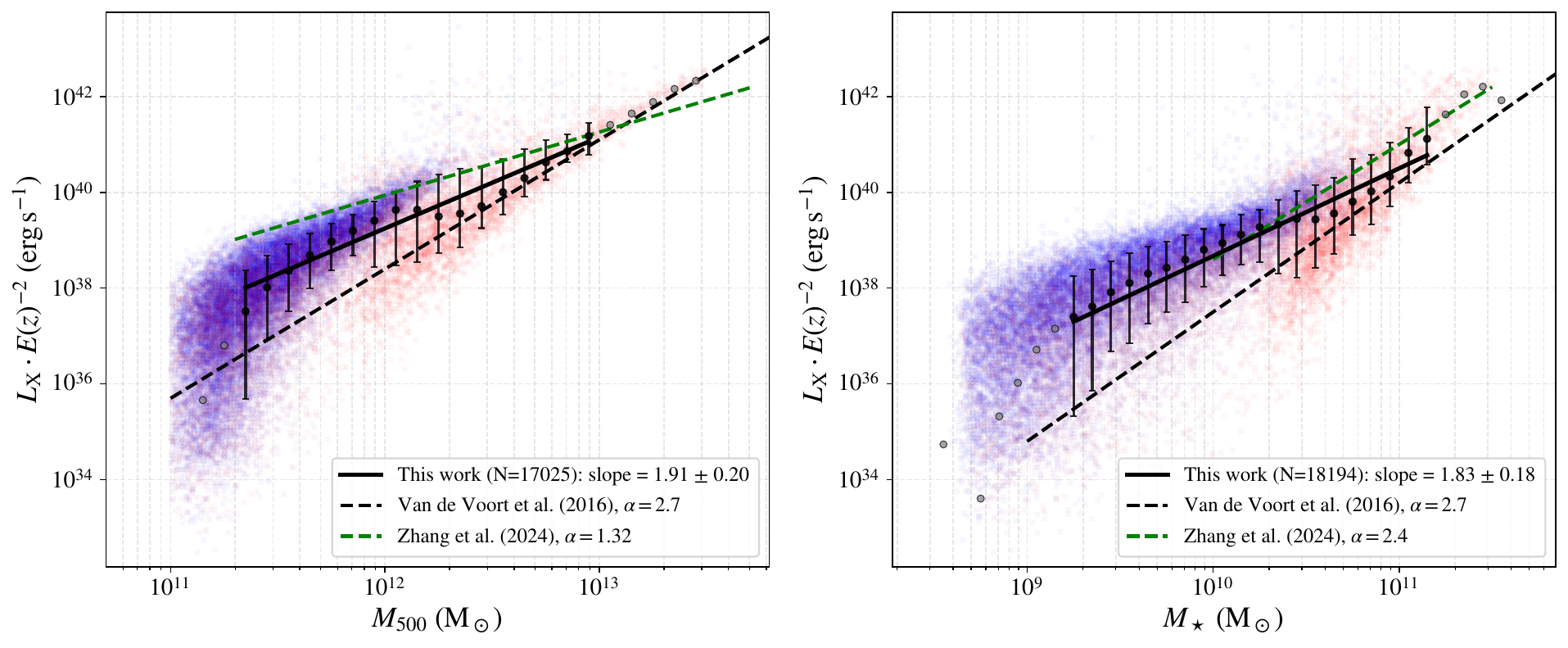}
    \caption{Comparison of the combined-sample $L_{\mathrm{X}}$--mass relations with published results. The sample pools halo--snapshot entries at $z=0$, $0.1$, and $0.2$, with colors following Figure~\ref{fig:LX_MassScaling}. Each luminosity is rescaled by $E(z)^{-2}$ before pooling and binning. Black points show the median rescaled luminosities, including zero values, in accepted 0.1 dex mass bins. Each accepted bin contains at least 100 entries and has a positive 16th-percentile luminosity. The bars show the 16th--84th percentiles of individual luminosities within each bin. Gray points show positive medians in bins excluded because they contain fewer than 100 entries or have a nonpositive 16th-percentile luminosity. The black solid line is the descriptive fit obtained with the prescription in Section~\ref{subsec:Scaling Relation with Mass: Fitting Method}. The displayed $N$ counts positive-luminosity halo--snapshot entries whose masses lie between the centers of the lowest- and highest-mass accepted bins. It is not the total number of entries contributing to the fitted bins. Black and green dashed curves show the relations of \citet{Voort2016MNRAS.463.4533V} and \citet{Zhang2024A&A...690A.268Z}, respectively. Differences in aperture, sample selection, and averaging remain, so these curves provide context rather than a like-for-like comparison. The $E(z)^{-2}$ luminosity rescaling is used only in this comparison.}
    \label{fig:LX_Mass_Comparison_with_previous_works}
\end{figure*}

For the alternative fits in Table~\ref{tab:scaling_geomean}, we compute the mean of $\log_{10}[L_{\mathrm{X}}/(\mathrm{erg\,s^{-1}})]$ for entries with $L_{\mathrm{X}}>0$ in each 0.1 dex mass bin and fit a straight line to these bin averages in log--log space. Exponentiating this mean gives the geometric-mean luminosity in $\mathrm{erg\,s^{-1}}$ for the positive-luminosity subset. Each accepted bin contains at least 100 entries with $L_{\mathrm{X}}>0$. These fits use the intrinsic luminosities without
the redshift rescaling applied in Figure~\ref{fig:LX_Mass_Comparison_with_previous_works}.

The lower and upper fitting widths are calculated from the logarithmic separations between the geometric mean and the 16th and 84th population percentiles of the positive-luminosity subset. Each width is constrained to be at least $10^{-4}$ dex. These widths are not standard errors of the mean logarithmic luminosity. The parameter estimates and quoted widths have
the same conditional interpretation as those in Table~\ref{tab:scaling_median}.

The geometric mean differs from our fiducial median, which includes zero-luminosity entries. These zero values belong to the defined simulation
measurement and are not observational nondetections. Excluding them changes the population being summarized. The comparison also changes the bin statistic and fitting weights, and different mass bins can satisfy
the acceptance criteria. Differences between the fitted slopes therefore
cannot be attributed to removing zeros alone, nor does zero exclusion necessarily steepen the relation. We report these fits as an alternative summary of the sample, not as a controlled test of any one methodological choice.

Figure~\ref{fig:LX_Mass_Comparison_with_previous_works} places our combined-sample median relations alongside published measurements.
Following the convention of \citet{Zhang2024A&A...690A.268Z}, we rescale each halo--snapshot luminosity as $L_{\mathrm{X}}/E(z)^2$, where $E(z)=H(z)/H_0$, before pooling the snapshots and computing the binned medians. We evaluate $E(z)$ with the adopted cosmology, using $z=0$, $0.1$, and $0.2$ for snapshots 99, 91, and 84, respectively.

This luminosity rescaling is restricted to Figure~\ref{fig:LX_Mass_Comparison_with_previous_works}, and the geometric-mean fits in Table~\ref{tab:scaling_geomean} use unrescaled luminosities.
The factor provides a common normalization convention, not a general correction for the evolution of soft X-ray luminosity.
It does not account for changes in the gas temperature distribution or abundances, and it does not match the aperture, averaging procedure, sample selection, or instrumental treatment of the different measurements.

The measurements in Figure~\ref{fig:LX_Mass_Comparison_with_previous_works} do not use the same aperture or averaging procedure. The eROSITA All‑Sky Survey (eRASS) stacks of \citet{Zhang2024A&A...690A.268Z} estimate an arithmetic mean of projected emission within $R_{500}$, with individually detected point sources masked. Our points are medians of intrinsic luminosity from hot gas in the spherical shell between $0.15R_{500}$ and $R_{500}$. A point-source mask is not equivalent to removing this entire central region, and projected emission can include gas outside our three-dimensional aperture.
\citet{Voort2016MNRAS.463.4533V} also use projected luminosities within $R_{500c}$, but summarize each simulated halo by its temporal median over $z=0$--0.5. Their stellar masses are measured within 20 proper kpc. They report slopes of 2.7 with both halo mass and stellar mass, over $10^{11}<M_{500c}/M_\odot<10^{13}$ and $10^9<M_\star/M_\odot<10^{11.5}$, respectively. These definitions differ from our pooled-bin medians and 30 kpc stellar aperture. We therefore use the plotted relations as context, without attributing differences in their slopes or normalizations to a particular physical mechanism.

\begin{table*}[!t]
\centering
\caption{Descriptive Fits to the Geometric-mean $L_{\mathrm{X}}$--Mass Relations \label{tab:scaling_geomean}}
\small 
\setlength{\tabcolsep}{6pt} 
\begin{tabular*}{\textwidth}{@{\extracolsep{\fill}} lccccc @{}}
\hline\hline
Population & Mass & Slope & Intercept & $\log_{10}(M/M_\odot)$ & $N_{\rm fit}$ \\
\hline
\textbf{Quiescent} & $M_{500}$ & $2.62 \pm 0.54$ & $7.23 \pm 6.76$ & $[11.85, 12.85]$ & 2991 \\
 & $M_\star$ & $3.42 \pm 0.99$ & $3.01 \pm 10.68$ & $[10.35, 11.15]$ & 3117 \\
\hline
\textbf{Star-forming} & $M_{500}$ & $3.09 \pm 0.51$ & $2.75 \pm 5.99$ & $[11.05, 12.15]$ & 11412 \\
 & $M_\star$ & $1.89 \pm 0.25$ & $20.24 \pm 2.48$ & $[8.65, 10.55]$ & 11568 \\
\hline
\textbf{Transition} & $M_{500}$ & $3.38 \pm 0.60$ & $-1.03 \pm 7.08$ & $[11.05, 12.25]$ & 10019 \\
 & $M_\star$ & $2.12 \pm 0.27$ & $17.40 \pm 2.67$ & $[8.65, 10.65]$ & 10212 \\
\hline
\textbf{All Galaxies} & $M_{500}$ & $2.10 \pm 0.23$ & $14.00 \pm 2.87$ & $[11.05, 12.95]$ & 25299 \\
 & $M_\star$ & $1.89 \pm 0.22$ & $19.82 \pm 2.24$ & $[8.65, 11.15]$ & 25788 \\
\hline
\end{tabular*}
\vspace{1ex}
\parbox{\textwidth}{\footnotesize \textbf{Note.} Fits to the geometric mean of intrinsic $L_{\mathrm{X}}>0$ luminosities in 0.1 dex mass bins containing at least 100 positive-luminosity entries. No $E(z)^{-2}$ rescaling is applied. The fitting widths are derived from population percentiles of the positive-luminosity subset, not from the standard error of the bin mean. Parameter estimates are medians of the sampled coefficient distributions, and the quoted $\pm$ values are half their 16th--84th percentile widths. These widths are conditional on the adopted fitting prescription and are not calibrated sampling uncertainties of the binned relation. The reported log-mass limits are the centers of the lowest- and highest-mass accepted bins. $N_{\rm fit}$ counts positive-luminosity halo--snapshot entries whose masses lie between these centers, inclusive. It is not an exact count of the entries in the accepted bins. The change from the fiducial median analysis affects the summarized population, bin statistic, fitting weights, and potentially the accepted mass interval. Differences in slope therefore cannot be assigned to zero exclusion alone.}
\end{table*}

\section{Hot-gas Profiles and Radial Cell Occupancy} \label{appendix: B}

\begin{figure*}[t!]
    \centering
    \includegraphics[width=\textwidth]{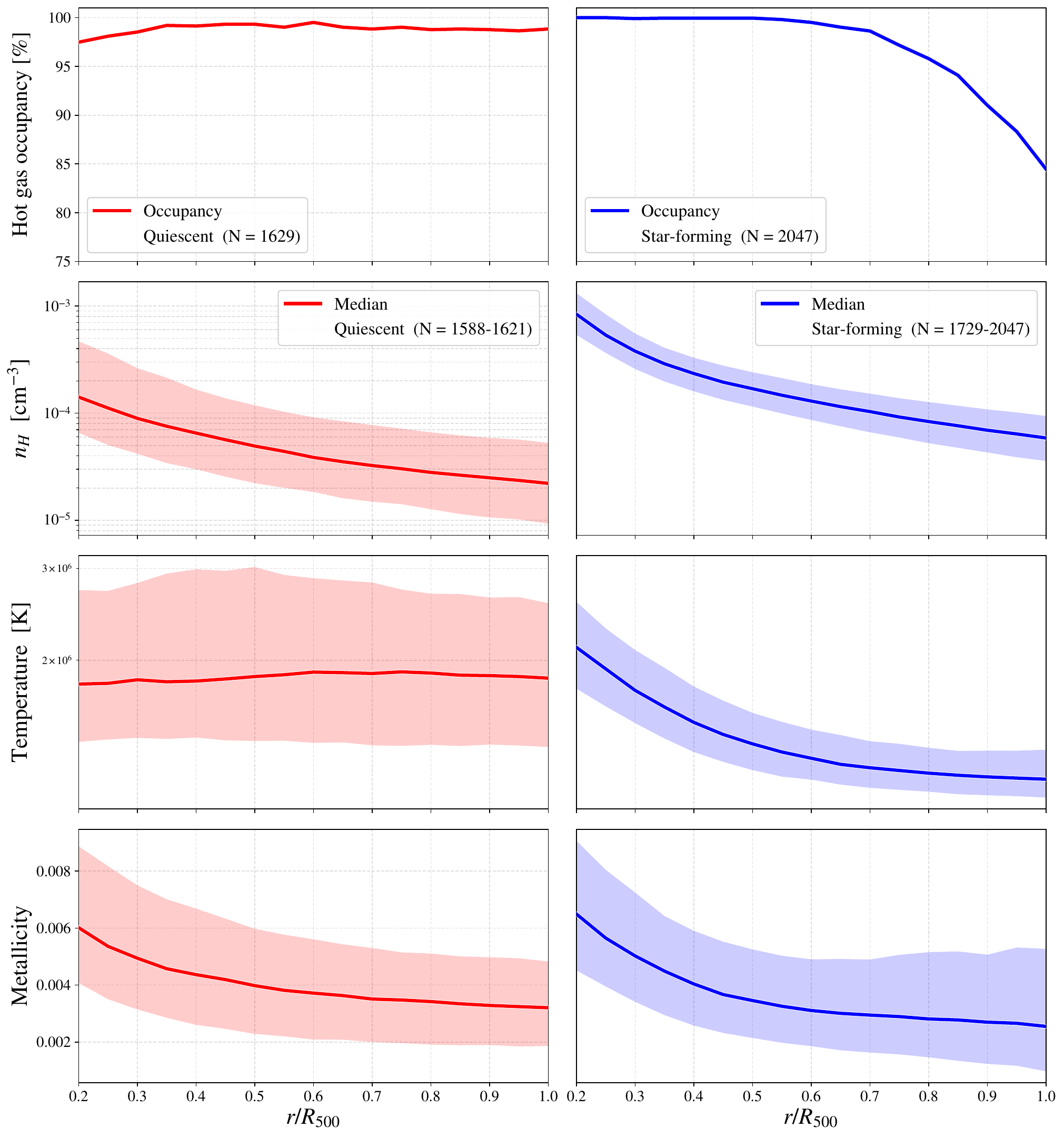}
    \caption{Hot-gas properties of the Milky Way scale sample pooled over $z=0$, $0.1$, and $0.2$. The horizontal coordinate marks the outer edge of each spherical shell in units of the three-dimensional radius $r/R_{500}$. The top row shows the fraction of halo--snapshot entries containing at least one cell with $T\geq10^6\,\mathrm{K}$ in each shell. The lower rows show the mass-weighted hydrogen number density, temperature, and metal mass fraction of this gas. Red and blue lines denote the quiescent and star-forming samples, respectively. In the lower rows, solid lines are medians across entries with hot gas in the relevant shell, and shading spans their 16th--84th percentiles rather than uncertainty on the median. Empty hot-gas shells contribute to the occupancy denominator but not to the property medians. The number of entries contributing to the lower rows therefore varies with radius.}
    \label{fig:RadialProfile_Gas_hot_Page}
\end{figure*}

Figure~\ref{fig:RadialProfile_Gas_hot_Page} shows the density, temperature, and metallicity of gas with $T\geq10^6\,\mathrm{K}$.
We use the same Milky Way scale sample and spherical shells as in
Section~\ref{sec:Star-Forming vs Quiescent galaxy in Milky Way mass galaxy}, pooling $z=0$, $0.1$, and $0.2$. For each halo and shell, we calculate the mass-weighted value of a cell property $q_i$ as
\begin{equation}
    \langle q\rangle_M
    =
    \frac{\sum_i m_iq_i}{\sum_i m_i}.
\end{equation}
The plotted median and 16th--84th percentile interval are then calculated across halo--snapshot entries containing hot gas in that shell. Empty hot-gas shells are omitted from these property statistics.

The density profiles are therefore mass-weighted means of cell densities, not gas masses divided by the geometric shell volumes. The radial coordinate marks the outer edge of each shell, and $Z$ denotes the metal mass fraction rather than a metallicity normalized to solar abundance. These are conditional profiles of the selected phase, not averages over all entries regardless of whether a shell contains hot gas.

Electron and hydrogen number densities are related by $n_{\mathrm{e}}=x_e n_{\mathrm{H}}$, where $x_e$ is the cell's electron abundance relative to hydrogen. We show $n_{\mathrm{H}}$ in the figure. The luminosity calculation uses the electron and hydrogen number densities of each individual gas cell.

The star-forming sample has a higher median hot-gas density, with the largest contrast in the inner shells and a smaller difference toward the outskirts. Its median temperature decreases from approximately $2.1\times10^6\,\mathrm{K}$ in the inner CGM to $1.2\times10^6\,\mathrm{K}$ near $R_{500}$. The quiescent profile is flatter, at approximately $(1.8$--$1.9)\times10^6\,\mathrm{K}$. These temperatures describe the gas above the adopted threshold, not the full CGM temperature distribution.

Hot-cell occupancy decreases more rapidly with radius in the star-forming sample than in the quiescent sample. This statistic measures the presence of resolved gas above the adopted temperature threshold in each shell. It does not by itself show that a halo has lost gas or determine the extent of the gas beyond $R_{500}$.

\bibliographystyle{aasjournalv7}
\bibliography{ms}

@ARTICLE{Tumlinson2017ARA&A..55..389T,
       author = {{Tumlinson}, Jason and {Peeples}, Molly S. and {Werk}, Jessica K.},
        title = "{The Circumgalactic Medium}",
      journal = {\araa},
         year = 2017,
        month = aug,
       volume = {55},
       number = {1},
        pages = {389-432},
          doi = {10.1146/annurev-astro-091916-055240},
archivePrefix = {arXiv},
       eprint = {1709.09180},
 primaryClass = {astro-ph.GA},
       adsurl = {https://ui.adsabs.harvard.edu/abs/2017ARA&A..55..389T}
}

@ARTICLE{Fukugita1998ApJ...503..518F,
       author = {{Fukugita}, M. and {Hogan}, C.~J. and {Peebles}, P.~J.~E.},
        title = "{The Cosmic Baryon Budget}",
      journal = {\apj},
         year = 1998,
        month = aug,
       volume = {503},
       number = {2},
        pages = {518-530},
          doi = {10.1086/306025},
archivePrefix = {arXiv},
       eprint = {astro-ph/9712020},
 primaryClass = {astro-ph},
       adsurl = {https://ui.adsabs.harvard.edu/abs/1998ApJ...503..518F}
}

@ARTICLE{Bregman2007ARA&A..45..221B,
       author = {{Bregman}, Joel N.},
        title = "{The Search for the Missing Baryons at Low Redshift}",
      journal = {\araa},
         year = 2007,
        month = sep,
       volume = {45},
       number = {1},
        pages = {221-259},
          doi = {10.1146/annurev.astro.45.051806.110619},
archivePrefix = {arXiv},
       eprint = {0706.1787},
 primaryClass = {astro-ph},
       adsurl = {https://ui.adsabs.harvard.edu/abs/2007ARA&A..45..221B}
}

@ARTICLE{Veilleux2005ARA&A..43..769V,
       author = {{Veilleux}, Sylvain and {Cecil}, Gerald and {Bland-Hawthorn}, Joss},
        title = "{Galactic Winds}",
      journal = {\araa},
         year = 2005,
        month = sep,
       volume = {43},
       number = {1},
        pages = {769-826},
          doi = {10.1146/annurev.astro.43.072103.150610},
archivePrefix = {arXiv},
       eprint = {astro-ph/0504435},
 primaryClass = {astro-ph},
       adsurl = {https://ui.adsabs.harvard.edu/abs/2005ARA&A..43..769V}
}

@ARTICLE{Keres2005MNRAS.363....2K,
       author = {{Kere{\v{s}}}, Du{\v{s}}an and {Katz}, Neal and {Weinberg}, David H. and {Dav{\'e}}, Romeel},
        title = "{How do galaxies get their gas?}",
      journal = {\mnras},
         year = 2005,
        month = oct,
       volume = {363},
       number = {1},
        pages = {2-28},
          doi = {10.1111/j.1365-2966.2005.09451.x},
archivePrefix = {arXiv},
       eprint = {astro-ph/0407095},
 primaryClass = {astro-ph},
       adsurl = {https://ui.adsabs.harvard.edu/abs/2005MNRAS.363....2K}
}

@ARTICLE{Spitzer1956ApJ...124...20S,
       author = {{Spitzer}, Jr., Lyman},
        title = "{On a Possible Interstellar Galactic Corona.}",
      journal = {\apj},
         year = 1956,
        month = jul,
       volume = {124},
        pages = {20},
          doi = {10.1086/146200},
       adsurl = {https://ui.adsabs.harvard.edu/abs/1956ApJ...124...20S}
}

@ARTICLE{Munch1961ApJ...133...11M,
       author = {{M{\"u}nch}, Guido and {Zirin}, Harold},
        title = "{Interstellar Matter at Large Distances from the Galactic Plane.}",
      journal = {\apj},
         year = 1961,
        month = jan,
       volume = {133},
        pages = {11},
          doi = {10.1086/146999},
       adsurl = {https://ui.adsabs.harvard.edu/abs/1961ApJ...133...11M}
}

@ARTICLE{Bahcall1969ApJ...156L..63B,
       author = {{Bahcall}, John N. and {Spitzer}, Jr., Lyman},
        title = "{Absorption Lines Produced by Galactic Halos}",
      journal = {\apjl},
         year = 1969,
        month = may,
       volume = {156},
        pages = {L63},
          doi = {10.1086/180350},
       adsurl = {https://ui.adsabs.harvard.edu/abs/1969ApJ...156L..63B}
}

@ARTICLE{Bergeron1986A&A...155L...8B,
       author = {{Bergeron}, J.},
        title = "{The MG II absorption system in the QSO PKS 2128-12 : a galaxy disc/halo with a radius of 65 kpc.}",
      journal = {\aap},
         year = 1986,
        month = jan,
       volume = {155},
        pages = {L8-L11},
       adsurl = {https://ui.adsabs.harvard.edu/abs/1986A&A...155L...8B}
}

@ARTICLE{Tumlinson2011Sci...334..948T,
       author = {{Tumlinson}, J. and {Thom}, C. and {Werk}, J.~K. and {Prochaska}, J.~X. and {Tripp}, T.~M. and {Weinberg}, D.~H. and {Peeples}, M.~S. and {O'Meara}, J.~M. and {Oppenheimer}, B.~D. and {Meiring}, J.~D. and {Katz}, N.~S. and {Dav{\'e}}, R. and {Ford}, A.~B. and {Sembach}, K.~R.},
        title = "{The Large, Oxygen-Rich Halos of Star-Forming Galaxies Are a Major Reservoir of Galactic Metals}",
      journal = {Science},
         year = 2011,
        month = nov,
       volume = {334},
       number = {6058},
        pages = {948},
          doi = {10.1126/science.1209840},
archivePrefix = {arXiv},
       eprint = {1111.3980},
 primaryClass = {astro-ph.CO},
       adsurl = {https://ui.adsabs.harvard.edu/abs/2011Sci...334..948T}
}

@ARTICLE{Werk2014ApJ...792....8W,
       author = {{Werk}, Jessica K. and {Prochaska}, J. Xavier and {Tumlinson}, Jason and {Peeples}, Molly S. and {Tripp}, Todd M. and {Fox}, Andrew J. and {Lehner}, Nicolas and {Thom}, Christopher and {O'Meara}, John M. and {Ford}, Amanda Brady and {Bordoloi}, Rongmon and {Katz}, Neal and {Tejos}, Nicolas and {Oppenheimer}, Benjamin D. and {Dav{\'e}}, Romeel and {Weinberg}, David H.},
        title = "{The COS-Halos Survey: Physical Conditions and Baryonic Mass in the Low-redshift Circumgalactic Medium}",
      journal = {\apj},
         year = 2014,
        month = sep,
       volume = {792},
       number = {1},
          eid = {8},
        pages = {8},
          doi = {10.1088/0004-637X/792/1/8},
archivePrefix = {arXiv},
       eprint = {1403.0947},
 primaryClass = {astro-ph.CO},
       adsurl = {https://ui.adsabs.harvard.edu/abs/2014ApJ...792....8W}
}

@ARTICLE{Prochaska2019MNRAS.485..648P,
       author = {{Prochaska}, J. Xavier and {Zheng}, Yong},
        title = "{Probing Galactic haloes with fast radio bursts}",
      journal = {\mnras},
         year = 2019,
        month = may,
       volume = {485},
       number = {1},
        pages = {648-665},
          doi = {10.1093/mnras/stz261},
archivePrefix = {arXiv},
       eprint = {1901.11051},
 primaryClass = {astro-ph.GA},
       adsurl = {https://ui.adsabs.harvard.edu/abs/2019MNRAS.485..648P}
}

@ARTICLE{Macquart2020Natur.581..391M,
       author = {{Macquart}, J.-P. and {Prochaska}, J.~X. and {McQuinn}, M. and {Bannister}, K.~W. and {Bhandari}, S. and {Day}, C.~K. and {Deller}, A.~T. and {Ekers}, R.~D. and {James}, C.~W. and {Marnoch}, L. and {Os{\l}owski}, S. and {Phillips}, C. and {Ryder}, S.~D. and {Scott}, D.~R. and {Shannon}, R.~M. and {Tejos}, N.},
        title = "{A census of baryons in the Universe from localized fast radio bursts}",
      journal = {\nat},
         year = 2020,
        month = may,
       volume = {581},
       number = {7809},
        pages = {391-395},
          doi = {10.1038/s41586-020-2300-2},
archivePrefix = {arXiv},
       eprint = {2005.13161},
 primaryClass = {astro-ph.CO},
       adsurl = {https://ui.adsabs.harvard.edu/abs/2020Natur.581..391M}
}

@ARTICLE{Anderson2015MNRAS.449.3806A,
       author = {{Anderson}, Michael E. and {Gaspari}, Massimo and {White}, Simon D.~M. and {Wang}, Wenting and {Dai}, Xinyu},
        title = "{Unifying X-ray scaling relations from galaxies to clusters}",
      journal = {\mnras},
         year = 2015,
        month = jun,
       volume = {449},
       number = {4},
        pages = {3806-3826},
          doi = {10.1093/mnras/stv437},
archivePrefix = {arXiv},
       eprint = {1409.6965},
 primaryClass = {astro-ph.CO},
       adsurl = {https://ui.adsabs.harvard.edu/abs/2015MNRAS.449.3806A}
}

@ARTICLE{Zhang2024A&A...690A.267Z,
       author = {{Zhang}, Yi and {Comparat}, Johan and {Ponti}, Gabriele and {Merloni}, Andrea and {Nandra}, Kirpal and {Haberl}, Frank and {Locatelli}, Nicola and {Zhang}, Xiaoyuan and {Sanders}, Jeremy and {Zheng}, Xueying and {Liu}, Ang and {Popesso}, Paola and {Liu}, Teng and {Truong}, Nhut and {Pillepich}, Annalisa and {Predehl}, Peter and {Salvato}, Mara and {Shreeram}, Soumya and {Yeung}, Michael C.~H. and {Ni}, Qingling},
        title = "{The hot circumgalactic medium in the eROSITA All-Sky Survey: I. X-ray surface brightness profiles}",
      journal = {\aap},
         year = 2024,
        month = oct,
       volume = {690},
          eid = {A267},
        pages = {A267},
          doi = {10.1051/0004-6361/202449412},
archivePrefix = {arXiv},
       eprint = {2401.17308},
 primaryClass = {astro-ph.GA},
       adsurl = {https://ui.adsabs.harvard.edu/abs/2024A&A...690A.267Z}
}

@ARTICLE{Hand2012PhRvL.109d1101H,
       author = {{Hand}, Nick and {Addison}, Graeme E. and {Aubourg}, Eric and {Battaglia}, Nick and {Battistelli}, Elia S. and {Bizyaev}, Dmitry and {Bond}, J. Richard and {Brewington}, Howard and {Brinkmann}, Jon and {Brown}, Benjamin R. and {Das}, Sudeep and {Dawson}, Kyle S. and {Devlin}, Mark J. and {Dunkley}, Joanna and {Dunner}, Rolando and {Eisenstein}, Daniel J. and {Fowler}, Joseph W. and {Gralla}, Megan B. and {Hajian}, Amir and {Halpern}, Mark and {Hilton}, Matt and {Hincks}, Adam D. and {Hlozek}, Ren{\'e}e and {Hughes}, John P. and {Infante}, Leopoldo and {Irwin}, Kent D. and {Kosowsky}, Arthur and {Lin}, Yen-Ting and {Malanushenko}, Elena and {Malanushenko}, Viktor and {Marriage}, Tobias A. and {Marsden}, Danica and {Menanteau}, Felipe and {Moodley}, Kavilan and {Niemack}, Michael D. and {Nolta}, Michael R. and {Oravetz}, Daniel and {Page}, Lyman A. and {Palanque-Delabrouille}, Nathalie and {Pan}, Kaike and {Reese}, Erik D. and {Schlegel}, David J. and {Schneider}, Donald P. and {Sehgal}, Neelima and {Shelden}, Alaina and {Sievers}, Jon and {Sif{\'o}n}, Crist{\'o}bal and {Simmons}, Audrey and {Snedden}, Stephanie and {Spergel}, David N. and {Staggs}, Suzanne T. and {Swetz}, Daniel S. and {Switzer}, Eric R. and {Trac}, Hy and {Weaver}, Benjamin A. and {Wollack}, Edward J. and {Yeche}, Christophe and {Zunckel}, Caroline},
        title = "{Evidence of Galaxy Cluster Motions with the Kinematic Sunyaev-Zel'dovich Effect}",
      journal = {\prl},
         year = 2012,
        month = jul,
       volume = {109},
       number = {4},
          eid = {041101},
        pages = {041101},
          doi = {10.1103/PhysRevLett.109.041101},
archivePrefix = {arXiv},
       eprint = {1203.4219},
 primaryClass = {astro-ph.CO},
       adsurl = {https://ui.adsabs.harvard.edu/abs/2012PhRvL.109d1101H}
}

@ARTICLE{Schaan2021PhRvD.103f3513S,
       author = {{Schaan}, Emmanuel and {Ferraro}, Simone and {Amodeo}, Stefania and {Battaglia}, Nicholas and {Aiola}, Simone and {Austermann}, Jason E. and {Beall}, James A. and {Bean}, Rachel and {Becker}, Daniel T. and {Bond}, Richard J. and {Calabrese}, Erminia and {Calafut}, Victoria and {Choi}, Steve K. and {Denison}, Edward V. and {Devlin}, Mark J. and {Duff}, Shannon M. and {Duivenvoorden}, Adriaan J. and {Dunkley}, Jo and {D{\"u}nner}, Rolando and {Gallardo}, Patricio A. and {Guan}, Yilun and {Han}, Dongwon and {Hill}, J. Colin and {Hilton}, Gene C. and {Hilton}, Matt and {Hlo{\v{z}}ek}, Ren{\'e}e and {Hubmayr}, Johannes and {Huffenberger}, Kevin M. and {Hughes}, John P. and {Koopman}, Brian J. and {MacInnis}, Amanda and {McMahon}, Jeff and {Madhavacheril}, Mathew S. and {Moodley}, Kavilan and {Mroczkowski}, Tony and {Naess}, Sigurd and {Nati}, Federico and {Newburgh}, Laura B. and {Niemack}, Michael D. and {Page}, Lyman A. and {Partridge}, Bruce and {Salatino}, Maria and {Sehgal}, Neelima and {Schillaci}, Alessandro and {Sif{\'o}n}, Crist{\'o}bal and {Smith}, Kendrick M. and {Spergel}, David N. and {Staggs}, Suzanne and {Storer}, Emilie R. and {Trac}, Hy and {Ullom}, Joel N. and {Van Lanen}, Jeff and {Vale}, Leila R. and {van Engelen}, Alexander and {Maga{\~n}a}, Mariana Vargas and {Vavagiakis}, Eve M. and {Wollack}, Edward J. and {Xu}, Zhilei and {Atacama Cosmology Telescope Collaboration}},
        title = "{Atacama Cosmology Telescope: Combined kinematic and thermal Sunyaev-Zel'dovich measurements from BOSS CMASS and LOWZ halos}",
      journal = {\prd},
         year = 2021,
        month = mar,
       volume = {103},
       number = {6},
          eid = {063513},
        pages = {063513},
          doi = {10.1103/PhysRevD.103.063513},
archivePrefix = {arXiv},
       eprint = {2009.05557},
 primaryClass = {astro-ph.CO},
       adsurl = {https://ui.adsabs.harvard.edu/abs/2021PhRvD.103f3513S}
}

@ARTICLE{Battaglia2019BAAS...51c.297B,
       author = {{Battaglia}, Nicholas and {Hill}, J. Colin and {Amodeo}, Stefania and {Bartlett}, James G. and {Basu}, Kaustuv and {Erler}, Jens and {Ferraro}, Simone and {Hernquist}, Lars and {Madhavacheril}, Mathew and {McQuinn}, Matthew and {Mroczkowski}, Tony and {Nagai}, Daisuke and {Schaan}, Emmanuel and {Somerville}, Rachel and {Sunyaev}, Rashid and {Vogelsberger}, Mark and {Werk}, Jessica},
        title = "{Probing Feedback in Galaxy Formation with Millimeter-wave Observations}",
      journal = {\baas},
         year = 2019,
        month = may,
       volume = {51},
       number = {3},
          eid = {297},
        pages = {297},
          doi = {10.48550/arXiv.1903.04647},
archivePrefix = {arXiv},
       eprint = {1903.04647},
 primaryClass = {astro-ph.CO},
       adsurl = {https://ui.adsabs.harvard.edu/abs/2019BAAS...51c.297B}
}

@ARTICLE{White1978MNRAS.183..341W,
       author = {{White}, S.~D.~M. and {Rees}, M.~J.},
        title = "{Core condensation in heavy halos: a two-stage theory for galaxy formation and clustering.}",
      journal = {\mnras},
         year = 1978,
        month = may,
       volume = {183},
        pages = {341-358},
          doi = {10.1093/mnras/183.3.341},
       adsurl = {https://ui.adsabs.harvard.edu/abs/1978MNRAS.183..341W}
}

@ARTICLE{Birnboim2003MNRAS.345..349B,
       author = {{Birnboim}, Yuval and {Dekel}, Avishai},
        title = "{Virial shocks in galactic haloes?}",
      journal = {\mnras},
         year = 2003,
        month = oct,
       volume = {345},
       number = {1},
        pages = {349-364},
          doi = {10.1046/j.1365-8711.2003.06955.x},
archivePrefix = {arXiv},
       eprint = {astro-ph/0302161},
 primaryClass = {astro-ph},
       adsurl = {https://ui.adsabs.harvard.edu/abs/2003MNRAS.345..349B}
}

@ARTICLE{Sunyaev1970Ap&SS...7....3S,
       author = {{Sunyaev}, R.~A. and {Zeldovich}, Ya. B.},
        title = "{Small-Scale Fluctuations of Relic Radiation}",
      journal = {\apss},
         year = 1970,
        month = apr,
       volume = {7},
       number = {1},
        pages = {3-19},
          doi = {10.1007/BF00653471},
       adsurl = {https://ui.adsabs.harvard.edu/abs/1970Ap&SS...7....3S}
}

@ARTICLE{Sunyaev1972CoASP...4..173S,
       author = {{Sunyaev}, R.~A. and {Zeldovich}, Ya. B.},
        title = "{The Observations of Relic Radiation as a Test of the Nature of X-Ray Radiation from the Clusters of Galaxies}",
      journal = {Comments on Astrophysics and Space Physics},
         year = 1972,
        month = nov,
       volume = {4},
        pages = {173},
       adsurl = {https://ui.adsabs.harvard.edu/abs/1972CoASP...4..173S}
}

@ARTICLE{Sunyaev1980ARA&A..18..537S,
       author = {{Sunyaev}, R.~A. and {Zeldovich}, Ia. B.},
        title = "{Microwave background radiation as a probe of the contemporary structure and history of the universe}",
      journal = {\araa},
         year = 1980,
        month = jan,
       volume = {18},
        pages = {537-560},
          doi = {10.1146/annurev.aa.18.090180.002541},
       adsurl = {https://ui.adsabs.harvard.edu/abs/1980ARA&A..18..537S}
}

@ARTICLE{Mroczkowski2019SSRv..215...17M,
       author = {{Mroczkowski}, Tony and {Nagai}, Daisuke and {Basu}, Kaustuv and {Chluba}, Jens and {Sayers}, Jack and {Adam}, R{\'e}mi and {Churazov}, Eugene and {Crites}, Abigail and {Di Mascolo}, Luca and {Eckert}, Dominique and {Macias-Perez}, Juan and {Mayet}, Fr{\'e}d{\'e}ric and {Perotto}, Laurence and {Pointecouteau}, Etienne and {Romero}, Charles and {Ruppin}, Florian and {Scannapieco}, Evan and {ZuHone}, John},
        title = "{Astrophysics with the Spatially and Spectrally Resolved Sunyaev-Zeldovich Effects. A Millimetre/Submillimetre Probe of the Warm and Hot Universe}",
      journal = {\ssr},
         year = 2019,
        month = feb,
       volume = {215},
       number = {1},
          eid = {17},
        pages = {17},
          doi = {10.1007/s11214-019-0581-2},
archivePrefix = {arXiv},
       eprint = {1811.02310},
 primaryClass = {astro-ph.CO},
       adsurl = {https://ui.adsabs.harvard.edu/abs/2019SSRv..215...17M}
}

@ARTICLE{Sarazin1986RvMP...58....1S,
       author = {{Sarazin}, Craig L.},
        title = "{X-ray emission from clusters of galaxies}",
      journal = {Reviews of Modern Physics},
         year = 1986,
        month = jan,
       volume = {58},
       number = {1},
        pages = {1-115},
          doi = {10.1103/RevModPhys.58.1},
       adsurl = {https://ui.adsabs.harvard.edu/abs/1986RvMP...58....1S}
}

@ARTICLE{Anderson2011ApJ...737...22A,
       author = {{Anderson}, Michael E. and {Bregman}, Joel N.},
        title = "{Detection of a Hot Gaseous Halo around the Giant Spiral Galaxy NGC 1961}",
      journal = {\apj},
         year = 2011,
        month = aug,
       volume = {737},
       number = {1},
          eid = {22},
        pages = {22},
          doi = {10.1088/0004-637X/737/1/22},
archivePrefix = {arXiv},
       eprint = {1105.4614},
 primaryClass = {astro-ph.CO},
       adsurl = {https://ui.adsabs.harvard.edu/abs/2011ApJ...737...22A}
}

@ARTICLE{Lovisari2021Univ....7..139L,
       author = {{Lovisari}, Lorenzo and {Ettori}, Stefano and {Gaspari}, Massimo and {Giles}, Paul A.},
        title = "{Scaling Properties of Galaxy Groups}",
      journal = {Universe},
         year = 2021,
        month = may,
       volume = {7},
       number = {5},
          eid = {139},
        pages = {139},
          doi = {10.3390/universe7050139},
archivePrefix = {arXiv},
       eprint = {2106.13256},
 primaryClass = {astro-ph.CO},
       adsurl = {https://ui.adsabs.harvard.edu/abs/2021Univ....7..139L}
}

@ARTICLE{Forman1985ApJ...293..102F,
       author = {{Forman}, W. and {Jones}, C. and {Tucker}, W.},
        title = "{Hot Coronae around Early-Type Galaxies}",
      journal = {\apj},
         year = 1985,
        month = jun,
       volume = {293},
        pages = {102},
          doi = {10.1086/163218},
       adsurl = {https://ui.adsabs.harvard.edu/abs/1985ApJ...293..102F}
}

@ARTICLE{Strickland2004ApJS..151..193S,
       author = {{Strickland}, David K. and {Heckman}, Timothy M. and {Colbert}, Edward J.~M. and {Hoopes}, Charles G. and {Weaver}, Kimberly A.},
        title = "{A High Spatial Resolution X-Ray and H{\ensuremath{\alpha}} Study of Hot Gas in the Halos of Star-forming Disk Galaxies. I. Spatial and Spectral Properties of the Diffuse X-Ray Emission}",
      journal = {\apjs},
         year = 2004,
        month = apr,
       volume = {151},
       number = {2},
        pages = {193-236},
          doi = {10.1086/382214},
archivePrefix = {arXiv},
       eprint = {astro-ph/0306592},
 primaryClass = {astro-ph},
       adsurl = {https://ui.adsabs.harvard.edu/abs/2004ApJS..151..193S}
}

@ARTICLE{Li2013MNRAS.428.2085L,
       author = {{Li}, Jiang-Tao and {Wang}, Q. Daniel},
        title = "{Chandra survey of nearby highly inclined disc galaxies - I. X-ray measurements of galactic coronae}",
      journal = {\mnras},
         year = 2013,
        month = jan,
       volume = {428},
       number = {3},
        pages = {2085-2108},
          doi = {10.1093/mnras/sts183},
archivePrefix = {arXiv},
       eprint = {1210.2997},
 primaryClass = {astro-ph.CO},
       adsurl = {https://ui.adsabs.harvard.edu/abs/2013MNRAS.428.2085L}
}

@ARTICLE{Junhan2022ApJ...926..179K,
       author = {{Kim}, Junhan and {Golwala}, Sunil and {Bartlett}, James G. and {Amodeo}, Stefania and {Battaglia}, Nicholas and {Benson}, Andrew J. and {Hill}, J. Colin and {Hopkins}, Philip F. and {Hummels}, Cameron B. and {Moser}, Emily and {Orr}, Matthew E.},
        title = "{Probing Hot Gas Components of the Circumgalactic Medium in Cosmological Simulations with the Thermal Sunyaev-Zel'dovich Effect}",
      journal = {\apj},
         year = 2022,
        month = feb,
       volume = {926},
       number = {2},
          eid = {179},
        pages = {179},
          doi = {10.3847/1538-4357/ac4750},
archivePrefix = {arXiv},
       eprint = {2110.15381},
 primaryClass = {astro-ph.GA},
       adsurl = {https://ui.adsabs.harvard.edu/abs/2022ApJ...926..179K}
}

@ARTICLE{Zhang2024A&A...690A.268Z,
       author = {{Zhang}, Yi and {Comparat}, Johan and {Ponti}, Gabriele and {Merloni}, Andrea and {Nandra}, Kirpal and {Haberl}, Frank and {Truong}, Nhut and {Pillepich}, Annalisa and {Locatelli}, Nicola and {Zhang}, Xiaoyuan and {Sanders}, Jeremy and {Zheng}, Xueying and {Liu}, Ang and {Popesso}, Paola and {Liu}, Teng and {Predehl}, Peter and {Salvato}, Mara and {Shreeram}, Soumya and {Yeung}, Michael C.~H.},
        title = "{The hot circumgalactic medium in the eROSITA All-Sky Survey: II. Scaling relations between X-ray luminosity and galaxies' mass}",
      journal = {\aap},
         year = 2024,
        month = oct,
       volume = {690},
          eid = {A268},
        pages = {A268},
          doi = {10.1051/0004-6361/202449413},
archivePrefix = {arXiv},
       eprint = {2401.17309},
 primaryClass = {astro-ph.GA},
       adsurl = {https://ui.adsabs.harvard.edu/abs/2024A&A...690A.268Z}
}

@ARTICLE{Zhang2025A&A...693A.197Z,
       author = {{Zhang}, Yi and {Comparat}, Johan and {Ponti}, Gabriele and {Merloni}, Andrea and {Nandra}, Kirpal and {Haberl}, Frank and {Truong}, Nhut and {Pillepich}, Annalisa and {Popesso}, Paola and {Locatelli}, Nicola and {Zhang}, Xiaoyuan and {Sanders}, Jeremy and {Zheng}, Xueying and {Liu}, Ang and {Liu}, Teng and {Predehl}, Peter and {Salvato}, Mara and {Bruggen}, Marcus and {Shreeram}, Soumya and {Yeung}, Michael C.~H.},
        title = "{The hot circumgalactic medium in the eROSITA All-Sky Survey: III. Star-forming and quiescent galaxies}",
      journal = {\aap},
         year = 2025,
        month = jan,
       volume = {693},
          eid = {A197},
        pages = {A197},
          doi = {10.1051/0004-6361/202452273},
archivePrefix = {arXiv},
       eprint = {2411.19945},
 primaryClass = {astro-ph.GA},
       adsurl = {https://ui.adsabs.harvard.edu/abs/2025A&A...693A.197Z}
}

@ARTICLE{Popesso2025A&A...704A.278P,
       author = {{Popesso}, P. and {Marini}, I. and {Dolag}, K. and {Lamer}, G. and {Csizi}, B. and {Biffi}, V. and {Robothan}, A. and {Bravo}, M. and {Biviano}, A. and {Vladutescu-Zopp}, S. and {Lovisari}, L. and {Ettori}, S. and {Angelinelli}, M. and {Driver}, S. and {Toptun}, V. and {Dev}, A. and {Mazengo}, D. and {Merloni}, A. and {Zhang}, Y. and {Comparat}, J. and {Ponti}, G. and {Mroczkowski}, T. and {Bulbul}, E.},
        title = "{Average X-ray properties of galaxy groups: From Milky Way-like halos to massive clusters}",
      journal = {\aap},
         year = 2025,
        month = dec,
       volume = {704},
          eid = {A278},
        pages = {A278},
          doi = {10.1051/0004-6361/202453255},
archivePrefix = {arXiv},
       eprint = {2411.17120},
 primaryClass = {astro-ph.GA},
       adsurl = {https://ui.adsabs.harvard.edu/abs/2025A&A...704A.278P}
}

@ARTICLE{Chadayammuri2022ApJ...936L..15C,
       author = {{Chadayammuri}, Urmila and {Bogd{\'a}n}, {\'A}kos and {Oppenheimer}, Benjamin D. and {Kraft}, Ralph P. and {Forman}, William R. and {Jones}, Christine},
        title = "{Testing Galaxy Feedback Models with Resolved X-Ray Profiles of the Hot Circumgalactic Medium}",
      journal = {\apjl},
         year = 2022,
        month = sep,
       volume = {936},
       number = {1},
          eid = {L15},
        pages = {L15},
          doi = {10.3847/2041-8213/ac8936},
archivePrefix = {arXiv},
       eprint = {2203.01356},
 primaryClass = {astro-ph.GA},
       adsurl = {https://ui.adsabs.harvard.edu/abs/2022ApJ...936L..15C}
}

@ARTICLE{Comparat2022A&A...666A.156C,
       author = {{Comparat}, Johan and {Truong}, Nhut and {Merloni}, Andrea and {Pillepich}, Annalisa and {Ponti}, Gabriele and {Driver}, Simon and {Bellstedt}, Sabine and {Liske}, Joe and {Aird}, James and {Br{\"u}ggen}, Marcus and {Bulbul}, Esra and {Davies}, Luke and {Villalba}, Justo Antonio Gonz{\'a}lez and {Georgakakis}, Antonis and {Haberl}, Frank and {Liu}, Teng and {Maitra}, Chandreyee and {Nandra}, Kirpal and {Popesso}, Paola and {Predehl}, Peter and {Robotham}, Aaron and {Salvato}, Mara and {Thorne}, Jessica E. and {Zhang}, Yi},
        title = "{The eROSITA Final Equatorial Depth Survey (eFEDS). X-ray emission around star-forming and quiescent galaxies at 0.05 < z < 0.3}",
      journal = {\aap},
         year = 2022,
        month = oct,
       volume = {666},
          eid = {A156},
        pages = {A156},
          doi = {10.1051/0004-6361/202243101},
archivePrefix = {arXiv},
       eprint = {2201.05169},
 primaryClass = {astro-ph.GA},
       adsurl = {https://ui.adsabs.harvard.edu/abs/2022A&A...666A.156C}
}

@ARTICLE{Voort2016MNRAS.463.4533V,
       author = {{van de Voort}, Freeke and {Quataert}, Eliot and {Hopkins}, Philip F. and {Faucher-Gigu{\`e}re}, Claude-Andr{\'e} and {Feldmann}, Robert and {Kere{\v{s}}}, Du{\v{s}}an and {Chan}, T.~K. and {Hafen}, Zachary},
        title = "{The impact of stellar feedback on hot gas in galaxy haloes: the Sunyaev-Zel'dovich effect and soft X-ray emission}",
      journal = {\mnras},
         year = 2016,
        month = dec,
       volume = {463},
       number = {4},
        pages = {4533-4544},
          doi = {10.1093/mnras/stw2322},
archivePrefix = {arXiv},
       eprint = {1604.01397},
 primaryClass = {astro-ph.GA},
       adsurl = {https://ui.adsabs.harvard.edu/abs/2016MNRAS.463.4533V}
}

@ARTICLE{Truong2020MNRAS.494..549T,
       author = {{Truong}, Nhut and {Pillepich}, Annalisa and {Werner}, Norbert and {Nelson}, Dylan and {Lakhchaura}, Kiran and {Weinberger}, Rainer and {Springel}, Volker and {Vogelsberger}, Mark and {Hernquist}, Lars},
        title = "{X-ray signatures of black hole feedback: hot galactic atmospheres in IllustrisTNG and X-ray observations}",
      journal = {\mnras},
         year = 2020,
        month = may,
       volume = {494},
       number = {1},
        pages = {549-570},
          doi = {10.1093/mnras/staa685},
archivePrefix = {arXiv},
       eprint = {1911.11165},
 primaryClass = {astro-ph.GA},
       adsurl = {https://ui.adsabs.harvard.edu/abs/2020MNRAS.494..549T}
}

@ARTICLE{Oppenheimer2020ApJ...893L..24O,
       author = {{Oppenheimer}, Benjamin D. and {Bogd{\'a}n}, {\'A}kos and {Crain}, Robert A. and {ZuHone}, John A. and {Forman}, William R. and {Schaye}, Joop and {Wijers}, Nastasha A. and {Davies}, Jonathan J. and {Jones}, Christine and {Kraft}, Ralph P. and {Ghirardini}, Vittorio},
        title = "{EAGLE and Illustris-TNG Predictions for Resolved eROSITA X-Ray Observations of the Circumgalactic Medium around Normal Galaxies}",
      journal = {\apjl},
         year = 2020,
        month = apr,
       volume = {893},
       number = {1},
          eid = {L24},
        pages = {L24},
          doi = {10.3847/2041-8213/ab846f},
archivePrefix = {arXiv},
       eprint = {2003.13889},
 primaryClass = {astro-ph.GA},
       adsurl = {https://ui.adsabs.harvard.edu/abs/2020ApJ...893L..24O}
}

@ARTICLE{planck2016A&A...594A..13P,
       author = {{Planck Collaboration} and {Ade}, P.~A.~R. and {Aghanim}, N. and {Arnaud}, M. and {Ashdown}, M. and {Aumont}, J. and {Baccigalupi}, C. and {Banday}, A.~J. and {Barreiro}, R.~B. and {Bartlett}, J.~G. and {Bartolo}, N. and {Battaner}, E. and {Battye}, R. and {Benabed}, K. and {Beno{\^\i}t}, A. and {Benoit-L{\'e}vy}, A. and {Bernard}, J.-P. and {Bersanelli}, M. and {Bielewicz}, P. and {Bock}, J.~J. and {Bonaldi}, A. and {Bonavera}, L. and {Bond}, J.~R. and {Borrill}, J. and {Bouchet}, F.~R. and {Boulanger}, F. and {Bucher}, M. and {Burigana}, C. and {Butler}, R.~C. and {Calabrese}, E. and {Cardoso}, J.-F. and {Catalano}, A. and {Challinor}, A. and {Chamballu}, A. and {Chary}, R.-R. and {Chiang}, H.~C. and {Chluba}, J. and {Christensen}, P.~R. and {Church}, S. and {Clements}, D.~L. and {Colombi}, S. and {Colombo}, L.~P.~L. and {Combet}, C. and {Coulais}, A. and {Crill}, B.~P. and {Curto}, A. and {Cuttaia}, F. and {Danese}, L. and {Davies}, R.~D. and {Davis}, R.~J. and {de Bernardis}, P. and {de Rosa}, A. and {de Zotti}, G. and {Delabrouille}, J. and {D{\'e}sert}, F.-X. and {Di Valentino}, E. and {Dickinson}, C. and {Diego}, J.~M. and {Dolag}, K. and {Dole}, H. and {Donzelli}, S. and {Dor{\'e}}, O. and {Douspis}, M. and {Ducout}, A. and {Dunkley}, J. and {Dupac}, X. and {Efstathiou}, G. and {Elsner}, F. and {En{\ss}lin}, T.~A. and {Eriksen}, H.~K. and {Farhang}, M. and {Fergusson}, J. and {Finelli}, F. and {Forni}, O. and {Frailis}, M. and {Fraisse}, A.~A. and {Franceschi}, E. and {Frejsel}, A. and {Galeotta}, S. and {Galli}, S. and {Ganga}, K. and {Gauthier}, C. and {Gerbino}, M. and {Ghosh}, T. and {Giard}, M. and {Giraud-H{\'e}raud}, Y. and {Giusarma}, E. and {Gjerl{\o}w}, E. and {Gonz{\'a}lez-Nuevo}, J. and {G{\'o}rski}, K.~M. and {Gratton}, S. and {Gregorio}, A. and {Gruppuso}, A. and {Gudmundsson}, J.~E. and {Hamann}, J. and {Hansen}, F.~K. and {Hanson}, D. and {Harrison}, D.~L. and {Helou}, G. and {Henrot-Versill{\'e}}, S. and {Hern{\'a}ndez-Monteagudo}, C. and {Herranz}, D. and {Hildebrandt}, S.~R. and {Hivon}, E. and {Hobson}, M. and {Holmes}, W.~A. and {Hornstrup}, A. and {Hovest}, W. and {Huang}, Z. and {Huffenberger}, K.~M. and {Hurier}, G. and {Jaffe}, A.~H. and {Jaffe}, T.~R. and {Jones}, W.~C. and {Juvela}, M. and {Keih{\"a}nen}, E. and {Keskitalo}, R. and {Kisner}, T.~S. and {Kneissl}, R. and {Knoche}, J. and {Knox}, L. and {Kunz}, M. and {Kurki-Suonio}, H. and {Lagache}, G. and {L{\"a}hteenm{\"a}ki}, A. and {Lamarre}, J.-M. and {Lasenby}, A. and {Lattanzi}, M. and {Lawrence}, C.~R. and {Leahy}, J.~P. and {Leonardi}, R. and {Lesgourgues}, J. and {Levrier}, F. and {Lewis}, A. and {Liguori}, M. and {Lilje}, P.~B. and {Linden-V{\o}rnle}, M. and {L{\'o}pez-Caniego}, M. and {Lubin}, P.~M. and {Mac{\'\i}as-P{\'e}rez}, J.~F. and {Maggio}, G. and {Maino}, D. and {Mandolesi}, N. and {Mangilli}, A. and {Marchini}, A. and {Maris}, M. and {Martin}, P.~G. and {Martinelli}, M. and {Mart{\'\i}nez-Gonz{\'a}lez}, E. and {Masi}, S. and {Matarrese}, S. and {McGehee}, P. and {Meinhold}, P.~R. and {Melchiorri}, A. and {Melin}, J.-B. and {Mendes}, L. and {Mennella}, A. and {Migliaccio}, M. and {Millea}, M. and {Mitra}, S. and {Miville-Desch{\^e}nes}, M.-A. and {Moneti}, A. and {Montier}, L. and {Morgante}, G. and {Mortlock}, D. and {Moss}, A. and {Munshi}, D. and {Murphy}, J.~A. and {Naselsky}, P. and {Nati}, F. and {Natoli}, P. and {Netterfield}, C.~B. and {N{\o}rgaard-Nielsen}, H.~U. and {Noviello}, F. and {Novikov}, D. and {Novikov}, I. and {Oxborrow}, C.~A. and {Paci}, F. and {Pagano}, L. and {Pajot}, F. and {Paladini}, R. and {Paoletti}, D. and {Partridge}, B. and {Pasian}, F. and {Patanchon}, G. and {Pearson}, T.~J. and {Perdereau}, O. and {Perotto}, L. and {Perrotta}, F. and {Pettorino}, V. and {Piacentini}, F. and {Piat}, M. and {Pierpaoli}, E. and {Pietrobon}, D. and {Plaszczynski}, S. and {Pointecouteau}, E. and {Polenta}, G. and {Popa}, L. and {Pratt}, G.~W. and {Pr{\'e}zeau}, G.},
        title = "{Planck 2015 results. XIII. Cosmological parameters}",
      journal = {\aap},
         year = 2016,
        month = sep,
       volume = {594},
          eid = {A13},
        pages = {A13},
          doi = {10.1051/0004-6361/201525830},
archivePrefix = {arXiv},
       eprint = {1502.01589},
 primaryClass = {astro-ph.CO},
       adsurl = {https://ui.adsabs.harvard.edu/abs/2016A&A...594A..13P}
}

@ARTICLE{Naiman2018MNRAS.477.1206N,
       author = {{Naiman}, Jill P. and {Pillepich}, Annalisa and {Springel}, Volker and {Ramirez-Ruiz}, Enrico and {Torrey}, Paul and {Vogelsberger}, Mark and {Pakmor}, R{\"u}diger and {Nelson}, Dylan and {Marinacci}, Federico and {Hernquist}, Lars and {Weinberger}, Rainer and {Genel}, Shy},
        title = "{First results from the IllustrisTNG simulations: a tale of two elements - chemical evolution of magnesium and europium}",
      journal = {\mnras},
         year = 2018,
        month = jun,
       volume = {477},
       number = {1},
        pages = {1206-1224},
          doi = {10.1093/mnras/sty618},
archivePrefix = {arXiv},
       eprint = {1707.03401},
 primaryClass = {astro-ph.GA},
       adsurl = {https://ui.adsabs.harvard.edu/abs/2018MNRAS.477.1206N}
}

@ARTICLE{Springel2018MNRAS.475..676S,
       author = {{Springel}, Volker and {Pakmor}, R{\"u}diger and {Pillepich}, Annalisa and {Weinberger}, Rainer and {Nelson}, Dylan and {Hernquist}, Lars and {Vogelsberger}, Mark and {Genel}, Shy and {Torrey}, Paul and {Marinacci}, Federico and {Naiman}, Jill},
        title = "{First results from the IllustrisTNG simulations: matter and galaxy clustering}",
      journal = {\mnras},
         year = 2018,
        month = mar,
       volume = {475},
       number = {1},
        pages = {676-698},
          doi = {10.1093/mnras/stx3304},
archivePrefix = {arXiv},
       eprint = {1707.03397},
 primaryClass = {astro-ph.GA},
       adsurl = {https://ui.adsabs.harvard.edu/abs/2018MNRAS.475..676S}
}

@ARTICLE{Pillepich2018MNRAS.475..648P,
       author = {{Pillepich}, Annalisa and {Nelson}, Dylan and {Hernquist}, Lars and {Springel}, Volker and {Pakmor}, R{\"u}diger and {Torrey}, Paul and {Weinberger}, Rainer and {Genel}, Shy and {Naiman}, Jill P. and {Marinacci}, Federico and {Vogelsberger}, Mark},
        title = "{First results from the IllustrisTNG simulations: the stellar mass content of groups and clusters of galaxies}",
      journal = {\mnras},
         year = 2018,
        month = mar,
       volume = {475},
       number = {1},
        pages = {648-675},
          doi = {10.1093/mnras/stx3112},
archivePrefix = {arXiv},
       eprint = {1707.03406},
 primaryClass = {astro-ph.GA},
       adsurl = {https://ui.adsabs.harvard.edu/abs/2018MNRAS.475..648P}
}

@ARTICLE{Nelson2018MNRAS.475..624N,
       author = {{Nelson}, Dylan and {Pillepich}, Annalisa and {Springel}, Volker and {Weinberger}, Rainer and {Hernquist}, Lars and {Pakmor}, R{\"u}diger and {Genel}, Shy and {Torrey}, Paul and {Vogelsberger}, Mark and {Kauffmann}, Guinevere and {Marinacci}, Federico and {Naiman}, Jill},
        title = "{First results from the IllustrisTNG simulations: the galaxy colour bimodality}",
      journal = {\mnras},
         year = 2018,
        month = mar,
       volume = {475},
       number = {1},
        pages = {624-647},
          doi = {10.1093/mnras/stx3040},
archivePrefix = {arXiv},
       eprint = {1707.03395},
 primaryClass = {astro-ph.GA},
       adsurl = {https://ui.adsabs.harvard.edu/abs/2018MNRAS.475..624N}
}

@ARTICLE{Marinacci2018MNRAS.480.5113M,
       author = {{Marinacci}, Federico and {Vogelsberger}, Mark and {Pakmor}, R{\"u}diger and {Torrey}, Paul and {Springel}, Volker and {Hernquist}, Lars and {Nelson}, Dylan and {Weinberger}, Rainer and {Pillepich}, Annalisa and {Naiman}, Jill and {Genel}, Shy},
        title = "{First results from the IllustrisTNG simulations: radio haloes and magnetic fields}",
      journal = {\mnras},
         year = 2018,
        month = nov,
       volume = {480},
       number = {4},
        pages = {5113-5139},
          doi = {10.1093/mnras/sty2206},
archivePrefix = {arXiv},
       eprint = {1707.03396},
 primaryClass = {astro-ph.CO},
       adsurl = {https://ui.adsabs.harvard.edu/abs/2018MNRAS.480.5113M}
}

@ARTICLE{Davis1985ApJ...292..371D,
       author = {{Davis}, M. and {Efstathiou}, G. and {Frenk}, C.~S. and {White}, S.~D.~M.},
        title = "{The evolution of large-scale structure in a universe dominated by cold dark matter}",
      journal = {\apj},
         year = 1985,
        month = may,
       volume = {292},
        pages = {371-394},
          doi = {10.1086/163168},
       adsurl = {https://ui.adsabs.harvard.edu/abs/1985ApJ...292..371D}
}

@ARTICLE{Springel2001MNRAS.328..726S,
       author = {{Springel}, Volker and {White}, Simon D.~M. and {Tormen}, Giuseppe and {Kauffmann}, Guinevere},
        title = "{Populating a cluster of galaxies - I. Results at z=0}",
      journal = {\mnras},
         year = 2001,
        month = dec,
       volume = {328},
       number = {3},
        pages = {726-750},
          doi = {10.1046/j.1365-8711.2001.04912.x},
archivePrefix = {arXiv},
       eprint = {astro-ph/0012055},
 primaryClass = {astro-ph},
       adsurl = {https://ui.adsabs.harvard.edu/abs/2001MNRAS.328..726S}
}

@ARTICLE{Donnari2019MNRAS.485.4817D,
       author = {{Donnari}, Martina and {Pillepich}, Annalisa and {Nelson}, Dylan and {Vogelsberger}, Mark and {Genel}, Shy and {Weinberger}, Rainer and {Marinacci}, Federico and {Springel}, Volker and {Hernquist}, Lars},
        title = "{The star formation activity of IllustrisTNG galaxies: main sequence, UVJ diagram, quenched fractions, and systematics}",
      journal = {\mnras},
         year = 2019,
        month = jun,
       volume = {485},
       number = {4},
        pages = {4817-4840},
          doi = {10.1093/mnras/stz712},
archivePrefix = {arXiv},
       eprint = {1812.07584},
 primaryClass = {astro-ph.GA},
       adsurl = {https://ui.adsabs.harvard.edu/abs/2019MNRAS.485.4817D}
}

@ARTICLE{Wetzel2013MNRAS.432..336W,
       author = {{Wetzel}, Andrew R. and {Tinker}, Jeremy L. and {Conroy}, Charlie and {van den Bosch}, Frank C.},
        title = "{Galaxy evolution in groups and clusters: satellite star formation histories and quenching time-scales in a hierarchical Universe}",
      journal = {\mnras},
         year = 2013,
        month = jun,
       volume = {432},
       number = {1},
        pages = {336-358},
          doi = {10.1093/mnras/stt469},
archivePrefix = {arXiv},
       eprint = {1206.3571},
 primaryClass = {astro-ph.CO},
       adsurl = {https://ui.adsabs.harvard.edu/abs/2013MNRAS.432..336W}
}

@ARTICLE{Smith2001ApJ...556L..91S,
       author = {{Smith}, Randall K. and {Brickhouse}, Nancy S. and {Liedahl}, Duane A. and {Raymond}, John C.},
        title = "{Collisional Plasma Models with APEC/APED: Emission-Line Diagnostics of Hydrogen-like and Helium-like Ions}",
      journal = {\apjl},
         year = 2001,
        month = aug,
       volume = {556},
       number = {2},
        pages = {L91-L95},
          doi = {10.1086/322992},
archivePrefix = {arXiv},
       eprint = {astro-ph/0106478},
 primaryClass = {astro-ph},
       adsurl = {https://ui.adsabs.harvard.edu/abs/2001ApJ...556L..91S}
}

@ARTICLE{Foster2020Atoms...8...49F,
       author = {{Foster}, Adam R. and {Heuer}, Keri},
        title = "{PyAtomDB: Extending the AtomDB Atomic Database to Model New Plasma Processes and Uncertainties}",
      journal = {Atoms},
         year = 2020,
        month = aug,
       volume = {8},
       number = {3},
        pages = {49},
          doi = {10.3390/atoms8030049},
       adsurl = {https://ui.adsabs.harvard.edu/abs/2020Atoms...8...49F}
}

@ARTICLE{Springel2010MNRAS.401..791S,
       author = {{Springel}, Volker},
        title = "{E pur si muove: Galilean-invariant cosmological hydrodynamical simulations on a moving mesh}",
      journal = {\mnras},
         year = 2010,
        month = jan,
       volume = {401},
       number = {2},
        pages = {791-851},
          doi = {10.1111/j.1365-2966.2009.15715.x},
archivePrefix = {arXiv},
       eprint = {0901.4107},
 primaryClass = {astro-ph.CO},
       adsurl = {https://ui.adsabs.harvard.edu/abs/2010MNRAS.401..791S}
}

@ARTICLE{1989GeCoA..53..197A,
       author = {{Anders}, E. and {Grevesse}, N.},
        title = "{Abundances of the elements: Meteoritic and solar}",
      journal = {\gca},
         year = 1989,
        month = jan,
       volume = {53},
       number = {1},
        pages = {197-214},
          doi = {10.1016/0016-7037(89)90286-X},
       adsurl = {https://ui.adsabs.harvard.edu/abs/1989GeCoA..53..197A}
}

@ARTICLE{Donnari2021MNRAS.506.4760D,
       author = {{Donnari}, Martina and {Pillepich}, Annalisa and {Nelson}, Dylan and {Marinacci}, Federico and {Vogelsberger}, Mark and {Hernquist}, Lars},
        title = "{Quenched fractions in the IllustrisTNG simulations: comparison with observations and other theoretical models}",
      journal = {\mnras},
         year = 2021,
        month = oct,
       volume = {506},
       number = {4},
        pages = {4760-4780},
          doi = {10.1093/mnras/stab1950},
archivePrefix = {arXiv},
       eprint = {2008.00004},
 primaryClass = {astro-ph.GA},
       adsurl = {https://ui.adsabs.harvard.edu/abs/2021MNRAS.506.4760D}
}

@ARTICLE{Foreman-Mackey2013PASP..125..306F,
       author = {{Foreman-Mackey}, Daniel and {Hogg}, David W. and {Lang}, Dustin and {Goodman}, Jonathan},
        title = "{emcee: The MCMC Hammer}",
      journal = {\pasp},
         year = 2013,
        month = mar,
       volume = {125},
       number = {925},
        pages = {306},
          doi = {10.1086/670067},
archivePrefix = {arXiv},
       eprint = {1202.3665},
 primaryClass = {astro-ph.IM},
       adsurl = {https://ui.adsabs.harvard.edu/abs/2013PASP..125..306F}
}

@ARTICLE{Goodman2010CAMCS...5...65G,
       author = {{Goodman}, Jonathan and {Weare}, Jonathan},
        title = "{Ensemble samplers with affine invariance}",
      journal = {Communications in Applied Mathematics and Computational Science},
         year = 2010,
        month = jan,
       volume = {5},
       number = {1},
        pages = {65-80},
          doi = {10.2140/camcos.2010.5.65},
       adsurl = {https://ui.adsabs.harvard.edu/abs/2010CAMCS...5...65G}
}

@ARTICLE{Rodriguez-Gomez2015MNRAS.449...49R,
       author = {{Rodriguez-Gomez}, Vicente and {Genel}, Shy and {Vogelsberger}, Mark and {Sijacki}, Debora and {Pillepich}, Annalisa and {Sales}, Laura V. and {Torrey}, Paul and {Snyder}, Greg and {Nelson}, Dylan and {Springel}, Volker and {Ma}, Chung-Pei and {Hernquist}, Lars},
        title = "{The merger rate of galaxies in the Illustris simulation: a comparison with observations and semi-empirical models}",
      journal = {\mnras},
         year = 2015,
        month = may,
       volume = {449},
       number = {1},
        pages = {49-64},
          doi = {10.1093/mnras/stv264},
archivePrefix = {arXiv},
       eprint = {1502.01339},
 primaryClass = {astro-ph.GA},
       adsurl = {https://ui.adsabs.harvard.edu/abs/2015MNRAS.449...49R}
}

@ARTICLE{Bland-Hawthorn2016ARA&A..54..529B,
       author = {{Bland-Hawthorn}, Joss and {Gerhard}, Ortwin},
        title = "{The Galaxy in Context: Structural, Kinematic, and Integrated Properties}",
      journal = {\araa},
         year = 2016,
        month = sep,
       volume = {54},
        pages = {529-596},
          doi = {10.1146/annurev-astro-081915-023441},
archivePrefix = {arXiv},
       eprint = {1602.07702},
 primaryClass = {astro-ph.GA},
       adsurl = {https://ui.adsabs.harvard.edu/abs/2016ARA&A..54..529B}
}

@ARTICLE{Hu2003ApJ...584..702H,
       author = {{Hu}, Wayne and {Kravtsov}, Andrey V.},
        title = "{Sample Variance Considerations for Cluster Surveys}",
      journal = {\apj},
         year = 2003,
        month = feb,
       volume = {584},
       number = {2},
        pages = {702-715},
          doi = {10.1086/345846},
archivePrefix = {arXiv},
       eprint = {astro-ph/0203169},
 primaryClass = {astro-ph},
       adsurl = {https://ui.adsabs.harvard.edu/abs/2003ApJ...584..702H}
}

@ARTICLE{Mandelbaum2016MNRAS.457.3200M,
       author = {{Mandelbaum}, Rachel and {Wang}, Wenting and {Zu}, Ying and {White}, Simon and {Henriques}, Bruno and {More}, Surhud},
        title = "{Strong bimodality in the host halo mass of central galaxies from galaxy-galaxy lensing}",
      journal = {\mnras},
         year = 2016,
        month = apr,
       volume = {457},
       number = {3},
        pages = {3200-3218},
          doi = {10.1093/mnras/stw188},
archivePrefix = {arXiv},
       eprint = {1509.06762},
 primaryClass = {astro-ph.GA},
       adsurl = {https://ui.adsabs.harvard.edu/abs/2016MNRAS.457.3200M}
}

@ARTICLE{Tinker2013ApJ...778...93T,
       author = {{Tinker}, Jeremy L. and {Leauthaud}, Alexie and {Bundy}, Kevin and {George}, Matthew R. and {Behroozi}, Peter and {Massey}, Richard and {Rhodes}, Jason and {Wechsler}, Risa H.},
        title = "{Evolution of the Stellar-to-dark Matter Relation: Separating Star-forming and Passive Galaxies from z = 1 to 0}",
      journal = {\apj},
         year = 2013,
        month = dec,
       volume = {778},
       number = {2},
          eid = {93},
        pages = {93},
          doi = {10.1088/0004-637X/778/2/93},
archivePrefix = {arXiv},
       eprint = {1308.2974},
 primaryClass = {astro-ph.CO},
       adsurl = {https://ui.adsabs.harvard.edu/abs/2013ApJ...778...93T}
}

@ARTICLE{Rodriguez-Puebla2015ApJ...799..130R,
       author = {{Rodr{\'\i}guez-Puebla}, Aldo and {Avila-Reese}, Vladimir and {Yang}, Xiaohu and {Foucaud}, Sebastien and {Drory}, Niv and {Jing}, Y.~P.},
        title = "{The Stellar-to-Halo Mass Relation of Local Galaxies Segregates by Color}",
      journal = {\apj},
         year = 2015,
        month = feb,
       volume = {799},
       number = {2},
          eid = {130},
        pages = {130},
          doi = {10.1088/0004-637X/799/2/130},
archivePrefix = {arXiv},
       eprint = {1408.5407},
 primaryClass = {astro-ph.GA},
       adsurl = {https://ui.adsabs.harvard.edu/abs/2015ApJ...799..130R}
}

@ARTICLE{Weinberger2017MNRAS.465.3291W,
       author = {{Weinberger}, Rainer and {Springel}, Volker and {Hernquist}, Lars and {Pillepich}, Annalisa and {Marinacci}, Federico and {Pakmor}, R{\"u}diger and {Nelson}, Dylan and {Genel}, Shy and {Vogelsberger}, Mark and {Naiman}, Jill and {Torrey}, Paul},
        title = "{Simulating galaxy formation with black hole driven thermal and kinetic feedback}",
      journal = {\mnras},
         year = 2017,
        month = mar,
       volume = {465},
       number = {3},
        pages = {3291-3308},
          doi = {10.1093/mnras/stw2944},
archivePrefix = {arXiv},
       eprint = {1607.03486},
 primaryClass = {astro-ph.GA},
       adsurl = {https://ui.adsabs.harvard.edu/abs/2017MNRAS.465.3291W}
}

@ARTICLE{2013A&A...558A..33A,
       author = {{Astropy Collaboration} and {Robitaille}, Thomas P. and {Tollerud}, Erik J. and {Greenfield}, Perry and {Droettboom}, Michael and {Bray}, Erik and {Aldcroft}, Tom and {Davis}, Matt and {Ginsburg}, Adam and {Price-Whelan}, Adrian M. and {Kerzendorf}, Wolfgang E. and {Conley}, Alexander and {Crighton}, Neil and {Barbary}, Kyle and {Muna}, Demitri and {Ferguson}, Henry and {Grollier}, Fr{\'e}d{\'e}ric and {Parikh}, Madhura M. and {Nair}, Prasanth H. and {Unther}, Hans M. and {Deil}, Christoph and {Woillez}, Julien and {Conseil}, Simon and {Kramer}, Roban and {Turner}, James E.~H. and {Singer}, Leo and {Fox}, Ryan and {Weaver}, Benjamin A. and {Zabalza}, Victor and {Edwards}, Zachary I. and {Azalee Bostroem}, K. and {Burke}, D.~J. and {Casey}, Andrew R. and {Crawford}, Steven M. and {Dencheva}, Nadia and {Ely}, Justin and {Jenness}, Tim and {Labrie}, Kathleen and {Lim}, Pey Lian and {Pierfederici}, Francesco and {Pontzen}, Andrew and {Ptak}, Andy and {Refsdal}, Brian and {Servillat}, Mathieu and {Streicher}, Ole},
        title = "{Astropy: A community Python package for astronomy}",
      journal = {\aap},
         year = 2013,
        month = oct,
       volume = {558},
          eid = {A33},
        pages = {A33},
          doi = {10.1051/0004-6361/201322068},
archivePrefix = {arXiv},
       eprint = {1307.6212},
 primaryClass = {astro-ph.IM},
       adsurl = {https://ui.adsabs.harvard.edu/abs/2013A&A...558A..33A}
}

@ARTICLE{2018AJ....156..123A,
       author = {{Astropy Collaboration} and {Price-Whelan}, A.~M. and {Sip{\H{o}}cz}, B.~M. and {G{\"u}nther}, H.~M. and {Lim}, P.~L. and {Crawford}, S.~M. and {Conseil}, S. and {Shupe}, D.~L. and {Craig}, M.~W. and {Dencheva}, N. and {Ginsburg}, A. and {VanderPlas}, J.~T. and {Bradley}, L.~D. and {P{\'e}rez-Su{\'a}rez}, D. and {de Val-Borro}, M. and {Aldcroft}, T.~L. and {Cruz}, K.~L. and {Robitaille}, T.~P. and {Tollerud}, E.~J. and {Ardelean}, C. and {Babej}, T. and {Bach}, Y.~P. and {Bachetti}, M. and {Bakanov}, A.~V. and {Bamford}, S.~P. and {Barentsen}, G. and {Barmby}, P. and {Baumbach}, A. and {Berry}, K.~L. and {Biscani}, F. and {Boquien}, M. and {Bostroem}, K.~A. and {Bouma}, L.~G. and {Brammer}, G.~B. and {Bray}, E.~M. and {Breytenbach}, H. and {Buddelmeijer}, H. and {Burke}, D.~J. and {Calderone}, G. and {Cano Rodr{\'\i}guez}, J.~L. and {Cara}, M. and {Cardoso}, J.~V.~M. and {Cheedella}, S. and {Copin}, Y. and {Corrales}, L. and {Crichton}, D. and {D'Avella}, D. and {Deil}, C. and {Depagne}, {\'E}. and {Dietrich}, J.~P. and {Donath}, A. and {Droettboom}, M. and {Earl}, N. and {Erben}, T. and {Fabbro}, S. and {Ferreira}, L.~A. and {Finethy}, T. and {Fox}, R.~T. and {Garrison}, L.~H. and {Gibbons}, S.~L.~J. and {Goldstein}, D.~A. and {Gommers}, R. and {Greco}, J.~P. and {Greenfield}, P. and {Groener}, A.~M. and {Grollier}, F. and {Hagen}, A. and {Hirst}, P. and {Homeier}, D. and {Horton}, A.~J. and {Hosseinzadeh}, G. and {Hu}, L. and {Hunkeler}, J.~S. and {Ivezi{\'c}}, {\v{Z}}. and {Jain}, A. and {Jenness}, T. and {Kanarek}, G. and {Kendrew}, S. and {Kern}, N.~S. and {Kerzendorf}, W.~E. and {Khvalko}, A. and {King}, J. and {Kirkby}, D. and {Kulkarni}, A.~M. and {Kumar}, A. and {Lee}, A. and {Lenz}, D. and {Littlefair}, S.~P. and {Ma}, Z. and {Macleod}, D.~M. and {Mastropietro}, M. and {McCully}, C. and {Montagnac}, S. and {Morris}, B.~M. and {Mueller}, M. and {Mumford}, S.~J. and {Muna}, D. and {Murphy}, N.~A. and {Nelson}, S. and {Nguyen}, G.~H. and {Ninan}, J.~P. and {N{\"o}the}, M. and {Ogaz}, S. and {Oh}, S. and {Parejko}, J.~K. and {Parley}, N. and {Pascual}, S. and {Patil}, R. and {Patil}, A.~A. and {Plunkett}, A.~L. and {Prochaska}, J.~X. and {Rastogi}, T. and {Reddy Janga}, V. and {Sabater}, J. and {Sakurikar}, P. and {Seifert}, M. and {Sherbert}, L.~E. and {Sherwood-Taylor}, H. and {Shih}, A.~Y. and {Sick}, J. and {Silbiger}, M.~T. and {Singanamalla}, S. and {Singer}, L.~P. and {Sladen}, P.~H. and {Sooley}, K.~A. and {Sornarajah}, S. and {Streicher}, O. and {Teuben}, P. and {Thomas}, S.~W. and {Tremblay}, G.~R. and {Turner}, J.~E.~H. and {Terr{\'o}n}, V. and {van Kerkwijk}, M.~H. and {de la Vega}, A. and {Watkins}, L.~L. and {Weaver}, B.~A. and {Whitmore}, J.~B. and {Woillez}, J. and {Zabalza}, V. and {Astropy Contributors}},
        title = "{The Astropy Project: Building an Open-science Project and Status of the v2.0 Core Package}",
      journal = {\aj},
         year = 2018,
        month = sep,
       volume = {156},
       number = {3},
          eid = {123},
        pages = {123},
          doi = {10.3847/1538-3881/aabc4f},
archivePrefix = {arXiv},
       eprint = {1801.02634},
 primaryClass = {astro-ph.IM},
       adsurl = {https://ui.adsabs.harvard.edu/abs/2018AJ....156..123A}
}

@ARTICLE{2022ApJ...935..167A,
       author = {{Astropy Collaboration} and {Price-Whelan}, Adrian M. and {Lim}, Pey Lian and {Earl}, Nicholas and {Starkman}, Nathaniel and {Bradley}, Larry and {Shupe}, David L. and {Patil}, Aarya A. and {Corrales}, Lia and {Brasseur}, C.~E. and {N{\"o}the}, Maximilian and {Donath}, Axel and {Tollerud}, Erik and {Morris}, Brett M. and {Ginsburg}, Adam and {Vaher}, Eero and {Weaver}, Benjamin A. and {Tocknell}, James and {Jamieson}, William and {van Kerkwijk}, Marten H. and {Robitaille}, Thomas P. and {Merry}, Bruce and {Bachetti}, Matteo and {G{\"u}nther}, H. Moritz and {Aldcroft}, Thomas L. and {Alvarado-Montes}, Jaime A. and {Archibald}, Anne M. and {B{\'o}di}, Attila and {Bapat}, Shreyas and {Barentsen}, Geert and {Baz{\'a}n}, Juanjo and {Biswas}, Manish and {Boquien}, M{\'e}d{\'e}ric and {Burke}, D.~J. and {Cara}, Daria and {Cara}, Mihai and {Conroy}, Kyle E. and {Conseil}, Simon and {Craig}, Matthew W. and {Cross}, Robert M. and {Cruz}, Kelle L. and {D'Eugenio}, Francesco and {Dencheva}, Nadia and {Devillepoix}, Hadrien A.~R. and {Dietrich}, J{\"o}rg P. and {Eigenbrot}, Arthur Davis and {Erben}, Thomas and {Ferreira}, Leonardo and {Foreman-Mackey}, Daniel and {Fox}, Ryan and {Freij}, Nabil and {Garg}, Suyog and {Geda}, Robel and {Glattly}, Lauren and {Gondhalekar}, Yash and {Gordon}, Karl D. and {Grant}, David and {Greenfield}, Perry and {Groener}, Austen M. and {Guest}, Steve and {Gurovich}, Sebastian and {Handberg}, Rasmus and {Hart}, Akeem and {Hatfield-Dodds}, Zac and {Homeier}, Derek and {Hosseinzadeh}, Griffin and {Jenness}, Tim and {Jones}, Craig K. and {Joseph}, Prajwel and {Kalmbach}, J. Bryce and {Karamehmetoglu}, Emir and {Ka{\l}uszy{\'n}ski}, Miko{\l}aj and {Kelley}, Michael S.~P. and {Kern}, Nicholas and {Kerzendorf}, Wolfgang E. and {Koch}, Eric W. and {Kulumani}, Shankar and {Lee}, Antony and {Ly}, Chun and {Ma}, Zhiyuan and {MacBride}, Conor and {Maljaars}, Jakob M. and {Muna}, Demitri and {Murphy}, N.~A. and {Norman}, Henrik and {O'Steen}, Richard and {Oman}, Kyle A. and {Pacifici}, Camilla and {Pascual}, Sergio and {Pascual-Granado}, J. and {Patil}, Rohit R. and {Perren}, Gabriel I. and {Pickering}, Timothy E. and {Rastogi}, Tanuj and {Roulston}, Benjamin R. and {Ryan}, Daniel F. and {Rykoff}, Eli S. and {Sabater}, Jose and {Sakurikar}, Parikshit and {Salgado}, Jes{\'u}s and {Sanghi}, Aniket and {Saunders}, Nicholas and {Savchenko}, Volodymyr and {Schwardt}, Ludwig and {Seifert-Eckert}, Michael and {Shih}, Albert Y. and {Jain}, Anany Shrey and {Shukla}, Gyanendra and {Sick}, Jonathan and {Simpson}, Chris and {Singanamalla}, Sudheesh and {Singer}, Leo P. and {Singhal}, Jaladh and {Sinha}, Manodeep and {Sip{\H{o}}cz}, Brigitta M. and {Spitler}, Lee R. and {Stansby}, David and {Streicher}, Ole and {{\v{S}}umak}, Jani and {Swinbank}, John D. and {Taranu}, Dan S. and {Tewary}, Nikita and {Tremblay}, Grant R. and {de Val-Borro}, Miguel and {Van Kooten}, Samuel J. and {Vasovi{\'c}}, Zlatan and {Verma}, Shresth and {de Miranda Cardoso}, Jos{\'e} Vin{\'\i}cius and {Williams}, Peter K.~G. and {Wilson}, Tom J. and {Winkel}, Benjamin and {Wood-Vasey}, W.~M. and {Xue}, Rui and {Yoachim}, Peter and {Zhang}, Chen and {Zonca}, Andrea and {Astropy Project Contributors}},
        title = "{The Astropy Project: Sustaining and Growing a Community-oriented Open-source Project and the Latest Major Release (v5.0) of the Core Package}",
      journal = {\apj},
         year = 2022,
        month = aug,
       volume = {935},
       number = {2},
          eid = {167},
        pages = {167},
          doi = {10.3847/1538-4357/ac7c74},
archivePrefix = {arXiv},
       eprint = {2206.14220},
 primaryClass = {astro-ph.IM},
       adsurl = {https://ui.adsabs.harvard.edu/abs/2022ApJ...935..167A}
}

@ARTICLE{2020Natur.585..357H,
       author = {{Harris}, Charles R. and {Millman}, K. Jarrod and {van der Walt}, St{\'e}fan J. and {Gommers}, Ralf and {Virtanen}, Pauli and {Cournapeau}, David and {Wieser}, Eric and {Taylor}, Julian and {Berg}, Sebastian and {Smith}, Nathaniel J. and {Kern}, Robert and {Picus}, Matti and {Hoyer}, Stephan and {van Kerkwijk}, Marten H. and {Brett}, Matthew and {Haldane}, Allan and {del R{\'\i}o}, Jaime Fern{\'a}ndez and {Wiebe}, Mark and {Peterson}, Pearu and {G{\'e}rard-Marchant}, Pierre and {Sheppard}, Kevin and {Reddy}, Tyler and {Weckesser}, Warren and {Abbasi}, Hameer and {Gohlke}, Christoph and {Oliphant}, Travis E.},
        title = "{Array programming with NumPy}",
      journal = {\nat},
         year = 2020,
        month = sep,
       volume = {585},
       number = {7825},
        pages = {357-362},
          doi = {10.1038/s41586-020-2649-2},
archivePrefix = {arXiv},
       eprint = {2006.10256},
 primaryClass = {cs.MS},
       adsurl = {https://ui.adsabs.harvard.edu/abs/2020Natur.585..357H}
}

@ARTICLE{2020NaMet..17..261V,
       author = {{Virtanen}, Pauli and {Gommers}, Ralf and {Oliphant}, Travis E. and {Haberland}, Matt and {Reddy}, Tyler and {Cournapeau}, David and {Burovski}, Evgeni and {Peterson}, Pearu and {Weckesser}, Warren and {Bright}, Jonathan and {van der Walt}, St{\'e}fan J. and {Brett}, Matthew and {Wilson}, Joshua and {Millman}, K. Jarrod and {Mayorov}, Nikolay and {Nelson}, Andrew R.~J. and {Jones}, Eric and {Kern}, Robert and {Larson}, Eric and {Carey}, C.~J. and {Polat}, {\.I}lhan and {Feng}, Yu and {Moore}, Eric W. and {VanderPlas}, Jake and {Laxalde}, Denis and {Perktold}, Josef and {Cimrman}, Robert and {Henriksen}, Ian and {Quintero}, E.~A. and {Harris}, Charles R. and {Archibald}, Anne M. and {Ribeiro}, Ant{\^o}nio H. and {Pedregosa}, Fabian and {van Mulbregt}, Paul and {SciPy 1.  0 Contributors}},
        title = "{SciPy 1.0: fundamental algorithms for scientific computing in Python}",
      journal = {Nature Methods},
         year = 2020,
        month = feb,
       volume = {17},
        pages = {261-272},
          doi = {10.1038/s41592-019-0686-2},
archivePrefix = {arXiv},
       eprint = {1907.10121},
 primaryClass = {cs.MS},
       adsurl = {https://ui.adsabs.harvard.edu/abs/2020NaMet..17..261V}
}

@ARTICLE{2007CSE.....9...90H,
       author = {{Hunter}, John D.},
        title = "{Matplotlib: A 2D Graphics Environment}",
      journal = {Computing in Science and Engineering},
         year = 2007,
        month = jan,
       volume = {9},
       number = {3},
        pages = {90-95},
          doi = {10.1109/MCSE.2007.55},
       adsurl = {https://ui.adsabs.harvard.edu/abs/2007CSE.....9...90H}
}

@ARTICLE{2023MNRAS.525.1976T,
       author = {{Truong}, Nhut and {Pillepich}, Annalisa and {Nelson}, Dylan and {Bogd{\'a}n}, {\'A}kos and {Schellenberger}, Gerrit and {Chakraborty}, Priyanka and {Forman}, William R. and {Kraft}, Ralph and {Markevitch}, Maxim and {Ogorzalek}, Anna and {Oppenheimer}, Benjamin D. and {Sarkar}, Arnab and {Veilleux}, Sylvain and {Vogelsberger}, Mark and {Wang}, Q. Daniel and {Werner}, Norbert and {Zhuravleva}, Irina and {Zuhone}, John},
        title = "{X-ray metal line emission from the hot circumgalactic medium: probing the effects of supermassive black hole feedback}",
      journal = {\mnras},
         year = 2023,
        month = oct,
       volume = {525},
       number = {2},
        pages = {1976-1997},
          doi = {10.1093/mnras/stad2216},
archivePrefix = {arXiv},
       eprint = {2307.01277},
 primaryClass = {astro-ph.GA},
       adsurl = {https://ui.adsabs.harvard.edu/abs/2023MNRAS.525.1976T}
}

@ARTICLE{2025ApJ...994...89G,
       author = {{Grayson}, Skylar and {Scannapieco}, Evan and {Comparat}, Johan and {ZuHone}, John A. and {Zhang}, Yi and {Shreeram}, Soumya and {Br{\"u}ggen}, Marcus and {Bulbul}, Esra},
        title = "{The Hot Circumgalactic Medium in Stacked X-Rays: Observations versus Simulations}",
      journal = {\apj},
         year = 2025,
        month = nov,
       volume = {994},
       number = {1},
          eid = {89},
        pages = {89},
          doi = {10.3847/1538-4357/ae100f},
archivePrefix = {arXiv},
       eprint = {2506.09123},
 primaryClass = {astro-ph.GA},
       adsurl = {https://ui.adsabs.harvard.edu/abs/2025ApJ...994...89G}
}

@ARTICLE{2025A&A...703A.137S,
       author = {{Shreeram}, Soumya and {Comparat}, Johan and {Merloni}, Andrea and {Ponti}, Gabriele and {Popesso}, Paola and {Zhang}, Yi and {Nandra}, Kirpal and {Salvato}, Mara and {Marini}, Ilaria and {Buchner}, Johannes and {Locatelli}, Nicola and {Igo}, Zsofi},
        title = "{Retrieving the hot circumgalactic medium physics from the X-ray radial profile from eROSITA with an IlustrisTNG-based forward model}",
      journal = {\aap},
         year = 2025,
        month = nov,
       volume = {703},
          eid = {A137},
        pages = {A137},
          doi = {10.1051/0004-6361/202554508},
archivePrefix = {arXiv},
       eprint = {2504.03840},
 primaryClass = {astro-ph.CO},
       adsurl = {https://ui.adsabs.harvard.edu/abs/2025A&A...703A.137S}
}

@ARTICLE{2026A&A...706A.102Z,
       author = {{Zhang}, Yi and {Shreeram}, Soumya and {Ponti}, Gabriele and {Comparat}, Johan and {Merloni}, Andrea and {Qu}, Zhijie and {Li}, Jiangtao and {Bregman}, Joel N. and {Fang}, Taotao},
        title = "{On the baryon budget in the X-ray-emitting circumgalactic medium of Milky Way-mass galaxies}",
      journal = {\aap},
         year = 2026,
        month = feb,
       volume = {706},
          eid = {A102},
        pages = {A102},
          doi = {10.1051/0004-6361/202556835},
archivePrefix = {arXiv},
       eprint = {2511.17313},
 primaryClass = {astro-ph.GA},
       adsurl = {https://ui.adsabs.harvard.edu/abs/2026A&A...706A.102Z}
}

@ARTICLE{2026arXiv260621495A,
       author = {{Ald{\'a}s}, Franklin and {G{\'o}mez}, Facundo A. and {Springel}, Volker and {Pakmor}, R{\"u}diger and {Saro}, Alex and {Castro}, Tiago},
        title = "{X-ray emission maps and scaling relations in IllustrisTNG and MillenniumTNG: Differences between cluster and group regimes}",
      journal = {arXiv e-prints},
         year = 2026,
        month = jun,
          eid = {arXiv:2606.21495},
        pages = {arXiv:2606.21495},
          doi = {10.48550/arXiv.2606.21495},
archivePrefix = {arXiv},
       eprint = {2606.21495},
 primaryClass = {astro-ph.GA},
       adsurl = {https://ui.adsabs.harvard.edu/abs/2026arXiv260621495A}
}

@ARTICLE{2026arXiv260116499H,
       author = {{He}, Lin and {Li}, Zhiyuan},
        title = "{Constraints on the Hot Circumgalactic Medium around Nearby L* Galaxies from SRG/eROSITA All Sky Survey}",
      journal = {arXiv e-prints},
         year = 2026,
        month = jan,
          eid = {arXiv:2601.16499},
        pages = {arXiv:2601.16499},
          doi = {10.48550/arXiv.2601.16499},
archivePrefix = {arXiv},
       eprint = {2601.16499},
 primaryClass = {astro-ph.GA},
       adsurl = {https://ui.adsabs.harvard.edu/abs/2026arXiv260116499H}
}

@ARTICLE{2019ComAC...6....2N,
       author = {{Nelson}, Dylan and {Springel}, Volker and {Pillepich}, Annalisa and {Rodriguez-Gomez}, Vicente and {Torrey}, Paul and {Genel}, Shy and {Vogelsberger}, Mark and {Pakmor}, Ruediger and {Marinacci}, Federico and {Weinberger}, Rainer and {Kelley}, Luke and {Lovell}, Mark and {Diemer}, Benedikt and {Hernquist}, Lars},
        title = "{The IllustrisTNG simulations: public data release}",
      journal = {Computational Astrophysics and Cosmology},
         year = 2019,
        month = may,
       volume = {6},
       number = {1},
          eid = {2},
        pages = {2},
          doi = {10.1186/s40668-019-0028-x},
archivePrefix = {arXiv},
       eprint = {1812.05609},
 primaryClass = {astro-ph.GA},
       adsurl = {https://ui.adsabs.harvard.edu/abs/2019ComAC...6....2N}
}

@ARTICLE{2025A&A...695A...2V,
       author = {{Vladutescu-Zopp}, S. and {Biffi}, V. and {Dolag}, K.},
        title = "{Radial X-ray profiles of simulated galaxies: Contributions from hot gas and X-ray binaries}",
      journal = {\aap},
         year = 2025,
        month = mar,
       volume = {695},
          eid = {A2},
        pages = {A2},
          doi = {10.1051/0004-6361/202450989},
archivePrefix = {arXiv},
       eprint = {2406.02686},
 primaryClass = {astro-ph.GA},
       adsurl = {https://ui.adsabs.harvard.edu/abs/2025A&A...695A...2V}
}

@ARTICLE{2025ApJ...993..125S,
       author = {{Silich}, Emily M. and {ZuHone}, John and {Bellomi}, Elena and {Hummels}, Cameron and {Oppenheimer}, Benjamin and {Hopkins}, Philip F. and {Lochhaas}, Cassandra and {Ponnada}, Sam B. and {Vikhlinin}, Alexey},
        title = "{X-Ray Emission Signatures of Galactic Feedback in the Hot Circumgalactic Medium: Predictions from Cosmological Hydrodynamical Simulations}",
      journal = {\apj},
         year = 2025,
        month = nov,
       volume = {993},
       number = {1},
          eid = {125},
        pages = {125},
          doi = {10.3847/1538-4357/ae08a3},
archivePrefix = {arXiv},
       eprint = {2506.17440},
 primaryClass = {astro-ph.GA},
       adsurl = {https://ui.adsabs.harvard.edu/abs/2025ApJ...993..125S}
}

@ARTICLE{2026ApJ..1004..184Z,
       author = {{Zhang}, Zhijie and {Zhang}, Xiaoxia and {Fang}, Taotao and {Li}, Hui and {Bryan}, Greg L. and {Marinacci}, Federico and {Torrey}, Paul and {Vogelsberger}, Mark and {Wang}, Junfeng and {Xu}, Haiguang and {Yu}, Qingzheng and {Yuan}, Feng},
        title = "{Probing the Hot Gaseous Halos of Milky Way─like Galaxies in the TNG50 Simulation}",
      journal = {\apj},
         year = 2026,
        month = jun,
       volume = {1004},
       number = {2},
          eid = {184},
        pages = {184},
          doi = {10.3847/1538-4357/ae6ef0},
archivePrefix = {arXiv},
       eprint = {2604.24494},
 primaryClass = {astro-ph.GA},
       adsurl = {https://ui.adsabs.harvard.edu/abs/2026ApJ..1004..184Z}
}

\end{document}